\documentclass[
superscriptaddress,
final,
reprint,
showkeys,
pre,
aps,
floatfix,
]{revtex4-2}

\usepackage{tcolorbox}
\tcbset{
  width=1\textwidth,
  standard jigsaw,
  opacityback=0,   
}
\usepackage{pgffor}
\usepackage{graphicx}
\usepackage[utf8]{inputenc}
\usepackage{amsmath,amssymb,booktabs,latexsym,MnSymbol,bm,makecell}
\usepackage[%
  colorlinks=true,
  urlcolor=blue,
  linkcolor=blue,
  citecolor=blue
]{hyperref}
\usepackage[final]{pdfpages}
\makeatletter
\AtBeginDocument{\let\LS@rot\@undefined}
\makeatother
\begin{document}

\title{Two-by-two ordinal patterns in art paintings}

\author{Mateus M. Tarozo} 
\author{Arthur A. B. Pessa} 
\affiliation{Departamento de F\'isica, Universidade Estadual de Maring\'a -- Maring\'a, PR 87020-900, Brazil}

\author{Luciano Zunino} 
\affiliation{Centro de Investigaciones \'Opticas (CONICET La Plata - CIC - UNLP), 1897 Gonnet, La Plata, Argentina}
\affiliation{Departamento de Ciencias B\'asicas, Facultad de Ingenier\'ia, Universidad Nacional de La Plata (UNLP), 1900 La Plata, Argentina}

\author{Osvaldo A. Rosso} 
\affiliation{Instituto de Física, Universidade Federal de Alagoas -- Maceió 57072-900, Brazil}

\author{Matja{\v z} Perc} 
\affiliation{Faculty of Natural Sciences and Mathematics, University of Maribor, Koro{\v s}ka cesta 160, 2000 Maribor, Slovenia}
\affiliation{Community Healthcare Center Dr. Adolf Drolc Maribor, Ulica talcev 9, 2000 Maribor, Slovenia}
\affiliation{Department of Physics, Kyung Hee University, 26 Kyungheedae-ro, Dongdaemun-gu, Seoul 02447, Republic of Korea}
\affiliation{Complexity Science Hub, Metternichgasse 8, 1030 Vienna, Austria}
\affiliation{University College, Korea University, 145 Anam-ro, Seongbuk-gu, Seoul 02841, Republic of Korea}

\author{Haroldo V. Ribeiro} 
\email{hvr@dfi.uem.br}
\affiliation{Departamento de F\'isica, Universidade Estadual de Maring\'a -- Maring\'a, PR 87020-900, Brazil}

\date{\today}


\begin{abstract}
Quantitative analysis of visual arts has recently expanded to encompass a more extensive array of artworks due to the availability of large-scale digitized art collections. Consistent with formal analyses by art historians, many of these studies highlight the significance of encoding spatial structures within artworks to enhance our understanding of visual arts. However, defining universally applicable, interpretable, and sufficiently simple units that capture the essence of paintings and their artistic styles remains challenging. Here we examine ordering patterns in pixel intensities within two-by-two partitions of images from nearly 140,000 paintings created over the past thousand years. These patterns, categorized into eleven types based on arguments of continuity and symmetry, are both universally applicable and detailed enough to correlate with low-level visual features of paintings. We uncover a universal distribution of these patterns, with consistent prevalence within groups, yet modulated across groups by a nontrivial interplay between pattern smoothness and the likelihood of identical pixel intensities. This finding provides a standardized metric for comparing paintings and styles, further establishing a scale to measure deviations from the average prevalence. Our research also shows that these simple patterns carry valuable information for identifying painting styles, though styles generally exhibit considerable variability in the prevalence of ordinal patterns. Moreover, shifts in the prevalence of these patterns reveal a trend in which artworks increasingly diverge from the average incidence over time; however, this evolution is neither smooth nor uniform, with substantial variability in pattern prevalence, particularly after the 1930s.
\end{abstract}

\maketitle

\begin{tcolorbox}
{\onecolumngrid
\noindent\textbf{Significance statement}
\\Spatial patterns play a crucial role in both aesthetic appreciation and in defining the characteristics of artworks and artistic movements. Despite advancements in image processing, encoding these patterns into simple quantitative representations remains challenging. Our study addresses this by analyzing the ordering of pixel intensities in small image partitions, allowing us to represent paintings by a set of 75 ordinal patterns grouped into 11 categories based on continuity and symmetry. These patterns capture key visual features -- sharp edges, well-defined or textured elements, and vertical or horizontal structures -- providing insights into the artistic style of paintings, highlighting their variability and the complex, nonhomogeneous nature of art evolution.}
\end{tcolorbox}
\vspace{3em}
\clearpage
\twocolumngrid
\section{Introduction}

The pioneering work of Birkhoff in 1933~\citep{birkhoff1933aesthetic}, where he formulated an aesthetic measure based on the ratio between order and complexity, is often viewed as one of the first contemporary attempts to define universal mathematical principles for evaluating artistic aesthetics. His ideas, however, have origins extending back to ancient Greece and medieval philosophers~\citep{beardsley1975aesthetics}. Despite this long historical path, the empirical characterization of artworks using physical science approaches is relatively recent. The work by Taylor \textit{et al.}~\citep{taylor1999fractal} in the late 1990s on the fractal nature of Pollock's drip paintings was pivotal in this regard and spurred a multitude of further studies~\citep{jones2006fractal, taylor2006fractal, taylor2007authenticating, pedram2008mona, jones2009drip, hughes2010quantification, shamir2012computer, de2016order, alvarez2016fractal, elsa2017topological}, which in turn contributed to the emergence of the field of quantitative study of visual arts~\citep{castrejon2003nasca, koch2010, montagner2016statistics}. However, it was not until recent years that these quantitative efforts reached a larger scale and expanded beyond a limited number of artworks from specific artists or artistic styles. This expansion was primarily driven by the recent availability of extensive digitized art collections, which not only facilitated practical applications but also started to contribute to deepening our understanding of the cultural and social aspects of art and how they possibly evolved over the centuries~\citep{perc2020beauty}. 

Research by Manovich and coauthors~\citep{manovich2015data, yazdani2017quantifying, ushizima2021cultural} has pioneered the analysis of large-scale datasets of paintings and other forms of visual art to quantify artistic evolution through metrics such as average brightness and saturation. Changes in the use of color and contrast over time were also investigated by Kim \textit{et al.}~\citep{kim2014large} and Lee~\textit{et al.}~\citep{lee2018heterogeneity}, who identified significant differences across historical periods and artistic styles. Art paintings were further explored using a physics-inspired approach based on permutation entropy and statistical complexity by Sigaki~\textit{et al.}~\citep{sigaki2018history}, revealing a temporal evolution marked by transitions aligned with major historical periods. This methodology was subsequently applied by Valensise \textit{et al.}~\citep{valensise2021entropy} to assess the visual complexity of memes on social media platforms. Additionally, Lee and coauthors~\citep{lee2020dissecting} dissected the compositional structure of landscape paintings, finding that from 1600 to 1850, artworks typically featured a primary horizontal partition with a secondary vertical division, a style that gradually fell into disuse, yielding a preference for dual horizontal partitions in 20th-century paintings. More recently, Karjus~\textit{et al.}~\citep{karjus2023compression} introduced a representation space termed compression ensembles, designed by calculating the normalized compression size of various transformations applied to original images, a method that proved effective in quantifying the complexity of a broad array of visual arts. In another recent study, Lee and coauthors~\citep{lee2024social} have used a large-scale dataset of contemporary paintings sold at auctions to demonstrate that visual features play only a marginal role in predicting artwork prices.

Previous research thus demonstrates that, although art is traditionally regarded as qualitative or metaphysical, quantitative analyses -- often inspired by physics -- are not only feasible but have already yielded significant insights into the nature of art and contributed to more practical applications. Notable examples include the use of wavelet transforms for dating paintings~\citep{jafarpour2009stylistic}, authenticating unknown or disputed artworks~\citep{johnson2008image, polatkan2009detection}, removing cradle artifacts~\citep{yin2014digital}, as well as other computational techniques for crack detection, digital inpainting of cracks~\citep{cornelis2013crack}, and separating overlapping images in X-ray scans of paintings~\citep{deligiannis2016multi, sabetsarvestani2019artificial}. Another prominent research avenue is the automatic classification of painting styles, which initially relied primarily on shallow learning approaches~\citep{jafarpour2009stylistic, shamir2010impressionism, arora2012towards, wu2013painting} but, inspired by the success of convolutional neural networks in image processing~\citep{krizhevsky2012imageNet, goodfellow2016deep}, has shifted to deep learning methods~\citep{zujovic2009classifying, tan2016ceci, mao2017deepart, elgammal2018shape, sandoval2019two}. These studies, particularly those employing wavelet transforms and convolutional neural networks, underscore the importance of encoding the spatial structures within paintings and other visual artworks to deepen our understanding of art. Spatial patterns are indeed essential for aesthetic appreciation~\citep{hekkert2006design, hubbard2018aesthetics} and are frequently used to qualitatively describe and distinguish the key characteristics of artistic movements. This focus on spatial structures is notably evident in the works of W\"olfflin, a seminal figure in art history's development of formal analysis~\citep{warnke1989heinrich}, who, in 1915, introduced five contrasting conceptual pairs to distinguish between Renaissance and Baroque art~\citep{wolfflin1950principles}. These celebrated pairs -- linear vs. painterly, plane vs. recession, closed vs. open form, multiplicity vs. unity, and clearness vs. unclearness -- primarily address spatial patterns and formal structures in paintings rather than color aspects~\citep{darst1983renaissance}. A somewhat similar emphasis on spatial patterns also appears in the more recent works by art historians such as Sypher~\citep{wylie1978four} and Davis~\citep{davis2018general}.

Nevertheless, defining individual elements as units within artworks remains a complex challenge, despite the substantial research conducted to date. These units must be simple enough to apply universally, yet sufficiently detailed to encapsulate the essence of the paintings and their artistic styles. To address this challenge, we analyze ordering patterns among pixel intensities within two-by-two partitions in painting images using a crowd-sourced dataset comprising almost 140,000 artworks spanning a timeline of nearly a thousand years. Our study identifies 75 unique ordinal patterns, which we categorize into 11 groups based on arguments of continuity and symmetry. These patterns are easily interpretable and their prevalence correlates with low-level visual characteristics of paintings, including the presence of sharp edges, well-defined or textured elements, and vertical or horizontal structures. We observe a universal pattern in the distribution of two-by-two ordinal patterns across paintings, characterized by an approximately constant prevalence within each group, yet modulated across groups by a nontrivial interplay between pattern smoothness and the likelihood of finding identical pixel intensities. The existence of a universal prevalence in the incidence of ordinal patterns thus allows us to define a standardized measure that not only compares paintings and styles but also sets up a metric informing how distant a painting or an artistic style is from the average prevalence. Our research shows that the prevalence of ordinal patterns carries information for identifying the artistic style of paintings, outperforming baseline classifiers and further highlighting the critical role of patterns associated with identical values within the two-by-two partitions. This analysis also shows that styles typically display large variability in the prevalence of ordinal patterns, reflecting both the limitations of representing artworks through these patterns and the heterogeneous quality of our crowd-sourced dataset, as well as the intrinsic ambiguity often present in attributing a unique style to artworks. Additionally, we quantify the temporal evolution of paintings in our dataset by examining shifts in the prevalence of ordinal patterns, revealing a general trend wherein artworks increasingly diverge from the average incidence. Notably, this evolution is neither smooth nor homogeneous, showing substantial variability in the prevalence of ordinal patterns, especially after the 1930s.

In what follows, we detail these results, beginning with the data presentation and the definition of our two-by-two patterns. This is accompanied by an interpretation of these patterns in relation to the low-level visual features of paintings. We then examine the prevalence and variability of these patterns across all paintings in our dataset as well as within specific artistic styles. We further explore the potential of quantifying the evolution of paintings in our dataset by investigating changes in the incidence and variability of ordinal patterns. Finally, we conclude by summarizing our findings, acknowledging limitations related to the heterogeneous quality of the dataset, historiographic biases, and the localized nature of our ordinal patterns. The Methods Section provides detailed information about the implementation of our approach.

\section{Results}
The dataset used in our research is the same as the one introduced in Sigaki \textit{et al.}~\citep{sigaki2018history}, comprising 137,364 paintings encompassing 154 styles from 2,391 artists and spanning a timeline of nearly a thousand years. These images were collected from Wikiart, a crowd-sourced, encyclopedic compilation of digital reproductions of artworks, covering primarily Western paintings. Given the dataset's reliance on heterogeneous sources, there is a substantial variation in the quality of digital reproductions and in the metadata associated with the artworks. These factors naturally constrain the generalizability and interpretation of the findings we shall present and discuss. Nevertheless, in the absence of large-scale, well-curated alternatives, the Wikiart dataset stands out as perhaps the best source for digital reproductions of paintings and for exploring new quantitative approaches to assessing artwork features.

As detailed in the Methods Section~\ref{methods:data}, image files for each painting were converted to grayscale using the standard luminance transformation. Consequently, each painting is represented by a matrix $A$ with $n_y$ rows (image height) and $n_x$ columns (image width), where each entry $a_{ij}$ denotes the pixel intensity of the $i$-th row and the $j$-th column. Building on the ordinal framework inaugurated by Bandt and Pompe~\citep{bandt2002permutation} and the recent classification of two-by-two ordinal patterns by Bandt and Wittfeld~\citep{bandt2023two}, we propose to characterize paintings by their distributions of ordinal patterns.

To describe our approach, let us consider a hypothetical three-by-three image
\begin{equation*}
A =
\begin{bmatrix}
4 & 2 & 9 \\
5 & 3 & 1 \\
2 & 6 & 3
\end{bmatrix}\,
\end{equation*}
and sample it using a two-by-two sliding partition that moves one pixel at a time, both horizontally and vertically, resulting in the four partitions:
\begin{equation*}
A_0 =
\begin{bmatrix}
4 & 2\\
5 & 3
\end{bmatrix},~~
A_1 =
\begin{bmatrix}
2 & 9\\
3 & 1
\end{bmatrix},~~
A_2 =
\begin{bmatrix}
5 & 3\\
2 & 6
\end{bmatrix},~\text{and}~~
A_3 =
\begin{bmatrix}
3 & 1\\
6 & 3
\end{bmatrix}.
\end{equation*}
Subsequently, we assign relative ranks to the pixel values within these partitions. Thus, representing the flattened elements of these partitions as $(a_0,a_1,a_2,a_3)$, the first partition $A_0=(4,2,5,3)$ is described by $[2031]$ because $a_0=4$ is the second-largest element (rank 2), $a_1=2$ is the smallest element (rank 0), $a_2=5$ is the largest element (rank 3), and $a_3=3$ is the third-largest element (rank 1). Similarly, $A_1$ corresponds to $[1320]$ and $A_2$ to $[2103]$. If a partition contains identical values, the same rank is attributed to them. Therefore, the partition $A_3$, where the number $3$ (rank 1) appears twice, corresponds to the pattern $[1021]$. We can identify up to $75$ distinct ordinal patterns: $24$ occur in partitions without identical values; $36$ arise when two partition elements are equal ($12$ patterns for each of the three possible ranks of the identical values); $8$ appear when three partition elements are identical ($4$ patterns for each of the two possible ranks of the identical values); $6$ are found when there are two pairs of identical values within a partition; and $1$ corresponds to a partition where all elements are identical.

In assessing all ordinal patterns of an image, we calculate their relative frequencies to define a probability distribution $P=\{p_i;~i=1,\dots,75\}$, where $p_i$ represents the relative frequency of the $i$-th pattern. Unlike Bandt and Wittfeld's approach~\citep{bandt2023two}, our method explicitly incorporates the occurrence of identical pixel intensities rather than relying on their positions within partitions to resolve rank ties (see Methods Section~\ref{methods:withouties}). Consequently, the total number of possible patterns expands from 24 to 75, and as we shall verify, these additional patterns provide essential information for a deeper understanding of paintings and artistic styles. 

\begin{figure*}[ht!]
  \centering
  \includegraphics[width=\textwidth, keepaspectratio]{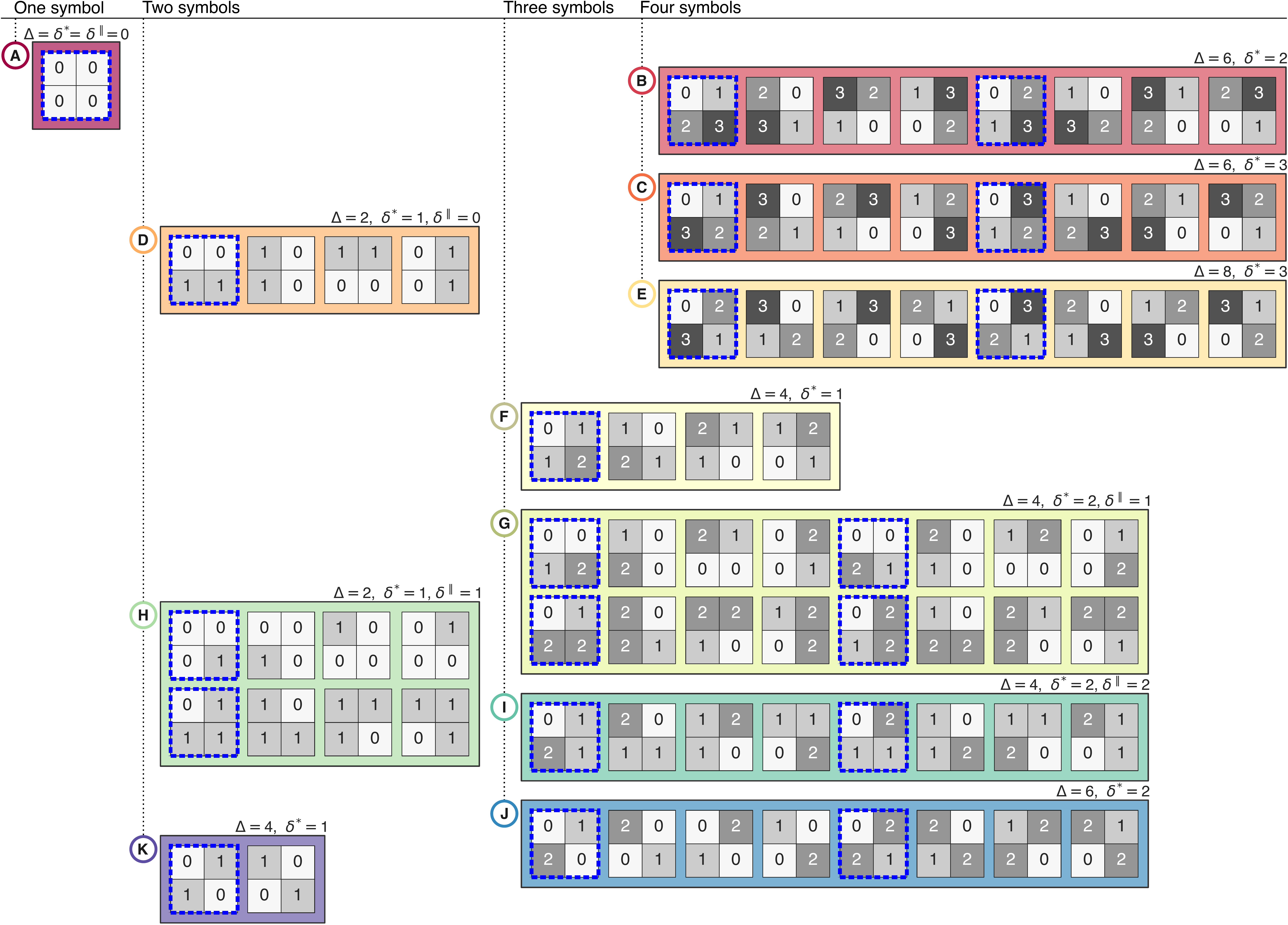}
  \caption{Overview of the 75 possible two-by-two ordinal patterns categorized into eleven groups based on their number of unique symbols and smoothness. Patterns are designated by capital letters from A to K and distinguished by different background colors. Smoothness is quantified by three metrics: sum of the absolute values of first-order differences among symbols ($\Delta$); maximum absolute value of these differences ($\delta^{*}$); and absolute value of the first-order difference between symbols aligned with either a row or column of identical symbols ($\delta^{\parallel}$). The pattern $[0000]$, the sole type A pattern, exhibits maximal smoothness with $\Delta=\delta^{*}=\delta^{\parallel}=0$ and corresponds to identical values in the associated partition. Groups D, H, and K, ranked by smoothness, contain patterns with two distinct symbols. Type D consists of four patterns with $\Delta=2$, $\delta^{*}=1$, and $\delta^{\parallel}=0$. Type H includes eight patterns with $\Delta=2$, $\delta^{*}=1$, and $\delta^{\parallel}=1$. Type K comprises two patterns with $\Delta=4$ and $\delta^{*}=1$. Patterns with three distinct symbols are categorized into four groups: F, G, I, and J, in order of smoothness. The four type F patterns are characterized by $\Delta=4$ and $\delta^{*}=1$. Type G, the largest group, includes sixteen patterns with $\Delta=4$, $\delta^{*}=2$, and $\delta^{\parallel}=1$. Type I comprises eight patterns, distinguished from type G by $\delta^{\parallel}=2$. Type J includes eight patterns with $\Delta=6$ and $\delta^{*}=2$. Finally, patterns with four distinct symbols are organized into groups B, C, and E, each containing eight patterns. Type B is the smoothest ($\Delta=6$ and $\delta^{*}=2$), followed by type C ($\Delta=6$ and $\delta^{*}=3$), and type E ($\Delta=8$ and $\delta^{*}=3$). Primary patterns within each group are highlighted with dashed blue lines. Rotations of primary patterns by quarter, half, and three-quarter turns generate all other patterns within each group.}
\label{fig:1}
\end{figure*}

We classify the identified patterns into eleven groups, designated by capital letters A to K. This classification is based on the number of distinct symbols each pattern exhibits and their smoothness degree, as depicted in Figure~\ref{fig:1}. We observe that certain patterns correspond to rotations of other patterns. Thus, we further identify a subset of patterns as ``primary patterns'' because their rotations by quarter, half, and three-quarter turns generate all other patterns within their respective groups. Each group contains one, two, or four primary patterns, which are highlighted in Figure~\ref{fig:1} by dashed blue lines. For a given pattern represented as $[x_1\, x_2\, x_3\,x_4]$, where each symbol $x_i$ (for $i=1,\dots,4$) takes a value from the set $\{0,1,2,3\}$, we quantify smoothness using three measures: sum of the absolute values of first-order differences among symbols ($\Delta=|x_1-x_2|+|x_3-x_4|+|x_1-x_3|+|x_2-x_4|$); maximum absolute value of these differences [$\delta^*=\max(|x_1-x_2|,|x_3-x_4|,|x_1-x_3|,|x_2-x_4|)$]; and absolute value of the first-order difference between symbols aligned along either a row or a column of identical symbols ($\delta^{\parallel}$). The values of $\delta^{\parallel}$ are thus only determined for patterns that feature two identical elements aligned along a row or column. For instance, the pattern $[0001]$ yields $\Delta = 2$, $\delta^* = 1$ and $\delta^{\parallel}=1$, whereas the pattern $[0110]$ results in $\Delta = 4$ and $\delta^* = 1$. We initially rank the smoothness of patterns based on $\Delta$; $\delta^*$ is subsequently used to resolve cases of equal $\Delta$ values, and $\delta^{\parallel}$ is used similarly when patterns display identical $\Delta$ and $\delta^*$ values.

The exclusive one-symbol pattern $[0000]$ is the sole member of group A. Characterized by its uniformity, this pattern exhibits the maximum smoothness with $\Delta=0$, indicative of identical pixel intensities within a partition. Two-symbol patterns occur in partitions that exhibit pairs or trios of identical values and are categorized into types D, H, and K. Among these, type $D$ represents the most homogeneous configuration ($\Delta=2$, $\delta^*=1$, and $\delta^{\parallel}=0$), typified by pairs of equal values aligned either in a column or a row with the primary pattern being $[0011]$. Type H, slightly less uniform, differs from type D by $\delta^{\parallel}=1$ and corresponds to trios of identical values that form corners within partitions whose primary patterns are $[0001]$ and $[0111]$. Type K, the least uniform within this group ($\Delta=4$), consists of patterns where pairs of equal values are aligned across both principal and secondary diagonals, as illustrated by the primary pattern $[0110]$. Three-symbol patterns emerge when partitions contain only a single pair of identical values; they are distributed across four groups: F, G, I, and J. Type F, with $[0112]$ as the primary pattern, is the smoothest within this category ($\Delta=4$ and $\delta^*=1$), displaying a configuration where the pair of identical values has the intermediate rank and is positioned diagonally, creating corner-like structures akin to type H. Type G is delineated by four primary types ($[0012]$, $[0021]$, $[0122]$, and $[0212]$) and type I by two ($[0121]$ and $[0211]$), both characterized by $\Delta=4$ and $\delta^*=2$ and corresponding to a configuration where the pair of identical values is located either along a column or a row. In Type G, the identical values assume the extremal ranks, whereas in Type I, they have the intermediate rank. This nuanced distinction results in a shift in $\delta^{\parallel}$ from $1$ to $2$, rendering Type I less uniform compared to Type G. Type J, identified by primary types $[0120]$ and $[0221]$, features the pair of identical values with the extremal ranks along the diagonals. The only difference between types F and J is the rank of their diagonal elements, but this variation alters the smoothness metric $\Delta$ from $4$ to $6$, categorizing type J as the least smooth among the three-symbol patterns.

Patterns with four symbols emerge when there are no identical values within a partition and comprise types B, C, and E. These are equivalent to the patterns named types I, II, and III by Bandt and Wittfeld~\citep{bandt2023two} (see Methods Section~\ref{methods:withouties}). Type B patterns, with $[0123]$ and $[0213]$ as primary patterns, occur when the partition values align in a three-dimensional plane, indicating simultaneous increases or decreases along both columns and rows. Type C, with primary patterns $[0132]$ and $[0312]$, displays $\Delta=6$ and $\delta^*=3$ and is less smooth than Type B ($\Delta=6$ and $\delta^*=2$). These patterns arise when pixel intensities within partitions simultaneously increase or decrease along one direction but exhibit distinct trends along the other. For instance, in $[0132]$, intensities in the first row increase from left to right, whereas those in the second decrease, with both columns increasing from top to bottom. Type E is the least smooth among four-symbol types and comprises the primary patterns $[0231]$ and $[0321]$. In these patterns, pixel intensities within partitions display distinct trends along both rows and columns. For example, in $[0231]$, the first row increases from left to right, while the second decreases and the first column increases from top to bottom while the second decreases. 

\begin{figure*}[ht!]
  \centering
  \includegraphics[width=\textwidth, keepaspectratio]{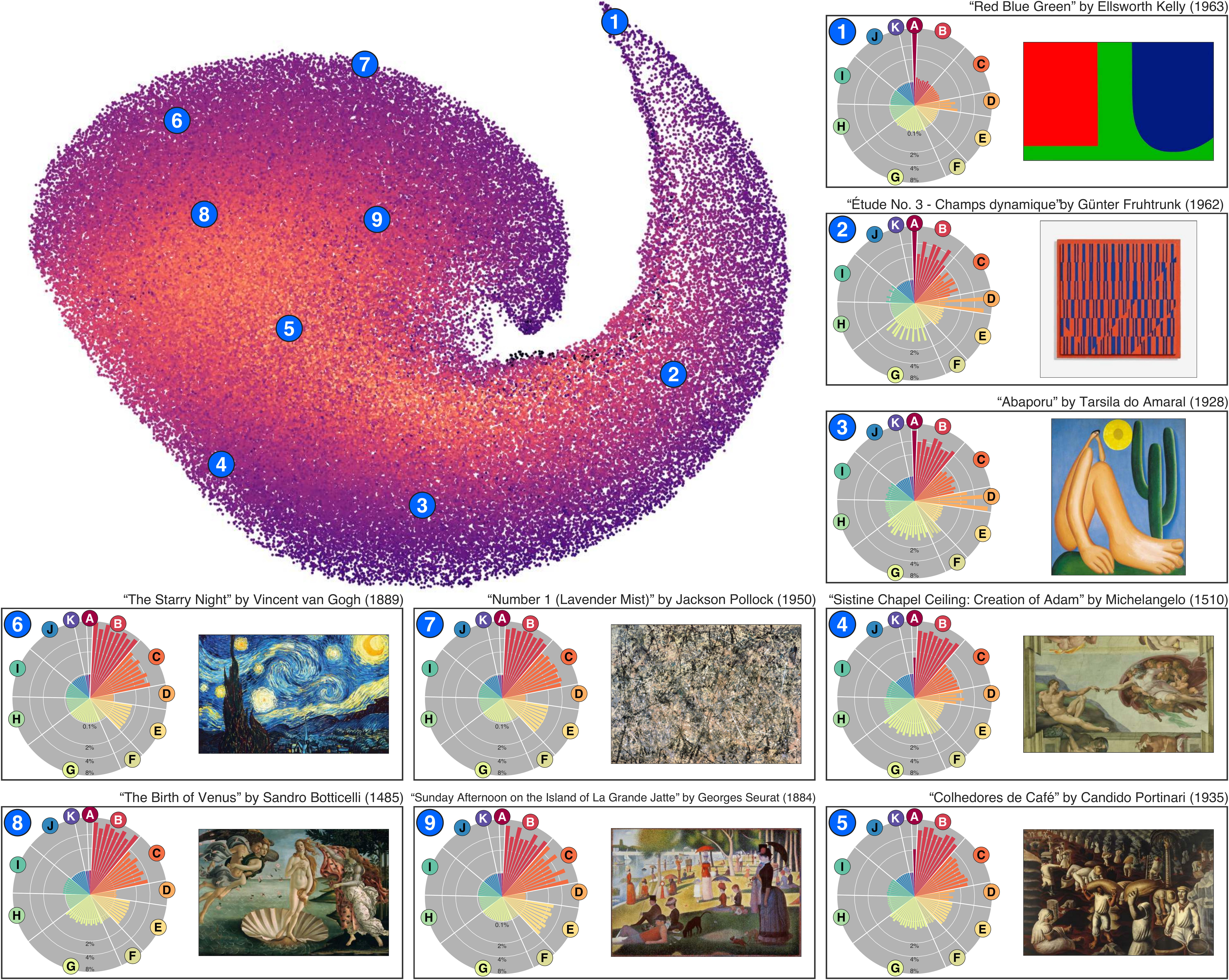}
  \caption{Representing paintings using two-by-two ordinal patterns. All paintings in our dataset are represented as 75-dimensional vectors, with each dimension corresponding to the relative frequency of a specific two-by-two ordinal pattern. The main panel displays a two-dimensional projection obtained from UMAP, where each dot corresponds to a painting. The numbers within blue circles indicate the positions of nine selected artworks. Surrounding panels illustrate the distribution of ordinal patterns, grouped by type, using circular bars on a logarithmic scale, accompanied by thumbnail images of the corresponding paintings. Paintings characterized by simple, well-defined design elements with sharp edges, such as the hard-edge painting by Ellsworth Kelly, typically exhibit a high frequency of the type A pattern. In contrast, paintings lacking sharp edges and well-defined elements, such as Jackson Pollock's drip paintings, display a negligible presence of this pattern. The ordinal patterns also reveal subtle nuances in artworks, such as the predominance of edges in specific directions, exemplified by the high frequency of two type D patterns in the Concretism work by G\"unter Fruhtrunk, or significant discontinuities among neighboring pixels, as observed in the pointillism artwork by Georges Seurat through the high frequency of two type E patterns.
  }
\label{fig:2}
\end{figure*}

We thus calculate the probability distribution of ordinal patterns $P=\{p_i;~i=1,\dots,75\}$ across all paintings in our dataset. In addition, we use our classification of ordinal patterns to arrange the elements of $P$ according to the type they belong from $A$ to $K$ and, within each type, we organize the probabilities $p_i$ as they appear in Figure~\ref{fig:1}. Consequently, $p_1$ corresponds to the probability of finding the pattern $[0000]$, $p_2$ is associated with $[0123]$, $p_3$ refers to $[2031]$, and so forth, up to $p_{75}$ that is related to $[1001]$. By doing so, we are mapping each image into a 75-dimensional vector, where each dimension represents the relative frequency of a specific two-by-two ordinal pattern. Although 75 dimensions certainly correspond to a high-dimensional space, it is significantly reduced compared to the original image resolution, which in our dataset typically exceeds one million pixels. Furthermore, these ordinal patterns all have straightforward interpretations and can be further categorized into 11 types. 

Figure~\ref{fig:2} presents a visualization of the probability distribution $P$ for all paintings using the uniform manifold approximation and projection (UMAP) dimensionality reduction algorithm~\citep{mcinnes2018umap} (see Methods Section~\ref{methods:UMAP}). The UMAP projects the 75-dimensional space of $P$ onto a plane, striving to preserve both local and global structures. In this projection, paintings or clusters of paintings that are proximal typically exhibit similar frequencies of ordinal patterns, while those that are distant display more dissimilar distributions. Notably, while UMAP preserves meaningful local structures~\citep{lause2024art}, it does not preserve high-dimensional distances~\citep{chari2023specious}; thus, insights derived from these projections remain primarily exploratory~\citep{lause2024art}. We annotate the positions of nine paintings, depicting their ordinal pattern distributions alongside images of the artworks. These distributions provide insights into various visual characteristics of the paintings. For instance, Ellsworth Kelly's ``Red Blue Green'' (1963) predominantly features type A and D patterns (with type A being much more frequent), reflecting the three sharply delineated colored regions of this abstract painting of the style Hard Edge Painting. We find a somewhat similar pattern in G\"unter Fruhtrunk's ``Étude No. 3 - Champs dynamique'' (1962) but with a significant presence of patterns $[1010]$ and $[0011]$, indicative of the painting's dominant vertical structures. Although more subtly, Tarsila do Amaral's ``Abaporu'' (1928) also reveals these vertical structures through a high occurrence of the same patterns. Conversely, artworks such as Jackson Pollock’s ``Number 1 (Lavender Mist)'' (1950), Vincent van Gogh's ``The Starry Night'' (1889), and Georges Seurat's ``Sunday Afternoon on the Island of La Grande Jatte'' (1884) practically do not exhibit the type A pattern. These ordinal patterns also capture more subtle nuances of artworks. For instance, Seurat's work is considered a notable example of the pointillist style of painting, in which the painter uses small brushstrokes to produce dots of locally distinct colors that, in turn, yield significant discontinuities among neighboring pixels. These discontinuities produce a relatively high frequency of two type E patterns. Furthermore, we observe a higher frequency of four type B patterns ($[0123]$, $[3210]$, $[1032]$, and $[2301]$) compared to the other four ($[2031]$, $[1302]$, $[0213]$, and $[3120]$) in Seurat's work. The patterns with high prevalence have smaller first-order differences horizontally than vertically, indicating that pixel intensities change more abruptly along the vertical direction and agreeing with the fact that Seurat started his artwork by applying a layer of small horizontal brushstrokes. Similar trends are observed in type C patterns, where patterns with greater first-order vertical differences along the vertical direction ($[0132]$, $[2310]$, $[1023]$, and $[3201]$) are more prevalent. 

It is important to acknowledge, however, that the frequency of ordinal patterns may be influenced by the resolution of the digital representation of the artwork, particularly for large paintings, where only high-resolution scans can adequately capture finer details. Seurat's ``Grande Jatte'', executed on a monumental canvas approximately 3 meters wide by 2 meters high, exemplifies this limitation, especially as we obtained a high-resolution representation of this artwork at 30,000$\times$19,970 pixels (Supplementary Figure~S1A). At this resolution, the small brushstrokes characteristic of this pointillist painting may emerge as well-defined regions, potentially increasing the prevalence of pattern types associated with identical pixel intensities, such as the type A pattern. To examine this effect, we analyze downscaled versions of the high-resolution image, starting from 200$\times$300 pixels (equivalent to sampling one pixel every 100 pixels along both the horizontal and vertical axes) up to the original resolution, incrementally increasing pixel sampling by one unit at each step and estimating the ordinal pattern distribution at each resolution. We observe that the prevalence of the type A pattern ([$0000$]) rises from 0 to 0.24\% as the pixel density increases, with similar behavior for pattern types D and H, and others associated with identical pixel intensities (Supplementary Figures~S1B and S1C). Despite the primary features of the ordinal distribution remaining relatively stable across resolutions (Supplementary Figure~S2), this illustrative case underscores the importance of caution when associating features of ordinal distributions with aesthetic patterns in artworks.

\begin{figure*}[ht!]
  \centering
  \includegraphics[width=\textwidth,keepaspectratio]{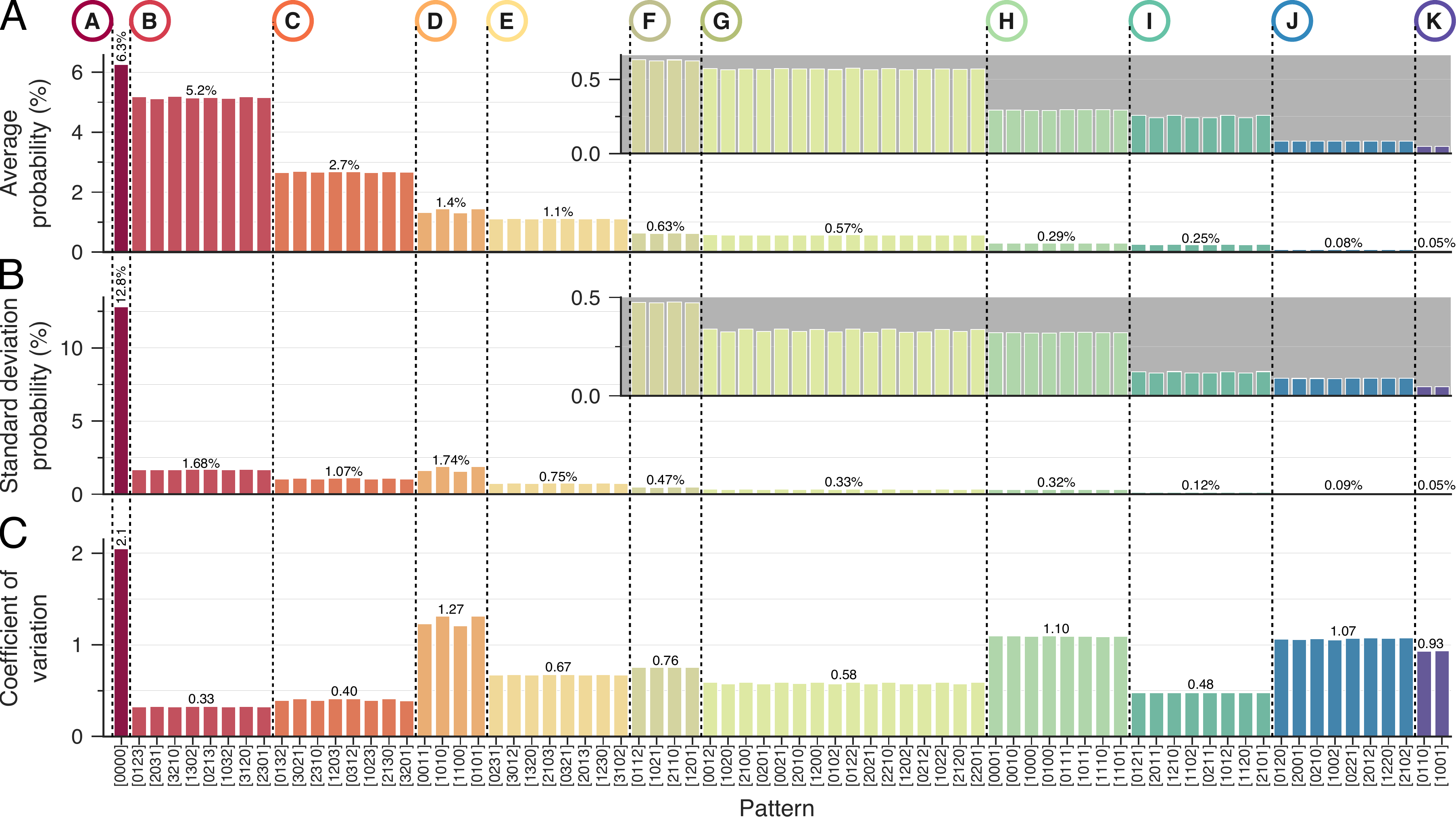}
  \caption{Universal patterns in the occurrence of two-by-two ordinal patterns. (A) Average probability of finding each of the 75 ordinal patterns across all images in our dataset. These patterns are categorized into eleven groups (A to K) and represented by different colors. Within each group, the probability of finding patterns remains approximately constant but decreases monotonically across alphabetically arranged groups. Within groups sharing the same number of distinct symbols, average probability diminishes as the roughness of the patterns increases. For example, type D patterns exhibit the greatest smoothness among types with two different symbols, followed by types H and K, with average probabilities of 1.4\%, 0.29\%, and 0.05\%, respectively. Across all types, the sequence of occurrence reflects a nontrivial interplay between pattern smoothness and the diminishing likelihood of repeating symbols. For instance, type F and K patterns, comparable in levels of smoothness, differ significantly in frequency; type K is less common due to the lower probability of encountering just two distinct symbols, compared to three in type F patterns. (B) Standard deviation and (C) coefficient of variation (standard deviation divided by the mean) for each ordinal pattern across all images in our dataset. The ordering of the standard deviation largely mirrors that of the average probabilities, except for type D patterns, which present the second-highest standard deviation. Conversely, the coefficient of variation across pattern types displays a distinct behavior, with types A, D, H, J, and K showing the highest coefficients of variation, respectively.
  }
\label{fig:3}
\end{figure*}

Considering the previous limitations, the examples highlighted in Figure~\ref{fig:2} illustrate the insights that can be gleaned from analyzing distributions of ordinal patterns. Integrating this analysis with domain-specific knowledge from the art field and high-quality digitalization of artworks offers a promising method for the individual examination of paintings. From a broader perspective, it raises the question of whether a universal prevalence of ordinal patterns exists in paintings. This query was also explored by Bandt and Wittfeld~\citep{bandt2023two} using their three types of ordinal patterns. In their study using a dataset of 25 natural textures, they observed that type I patterns are more prevalent than type II, which in turn surpasses type III in prevalence. This finding indicates that the prevalence of patterns correlates with their degree of smoothness, with smoother patterns appearing more frequently in natural textures. Despite the limited scale and variability of their dataset, it is plausible to assume that the smoothness of our eleven pattern types influences their occurrence. Indeed, that is the primary rationale for using smoothness to categorize our types. However, it is reasonable to suppose that the number of identical values within a partition also impacts the prevalence of corresponding patterns. To address this question, we calculate the average prevalence of the 75 ordinal patterns across all images in our dataset. Figure~\ref{fig:3}A reveals a clear hierarchical organization in the prevalence of ordinal patterns. Their average probabilities monotonically decrease across alphabetically arranged types and patterns within each group exhibit nearly identical prevalence. The smoothest pattern, type A, is also the most frequent, whereas the less frequent patterns belong to group K. Within groups sharing the same number of symbols (identical values), the average probability monotonically decreases as the roughness of their patterns increases. For instance, among patterns with two symbols, type D is more prevalent than type H, which is more prevalent than type K, following the same sequence in smoothness. This behavior is consistent for patterns with three and four symbols, ranked by prevalence and smoothness as (F, G, I, J) and (B, C, E), respectively. Conversely, no clear relationship between smoothness and prevalence exists across types with different numbers of symbols. Therefore, the prevalence of pattern types reflects a nontrivial interplay between their smoothness and the reduced likelihood of encountering identical values in two-by-two partitions.

We further explore the variability in the prevalence of ordinal patterns. Figures~\ref{fig:3}B and \ref{fig:3}C depict the standard deviation and the coefficient of variation (the ratio of standard deviation to mean) of the probability of each pattern across all images. The ranking by standard deviation largely corresponds to that by average prevalence, except for type D patterns, which exhibit the second-largest standard deviation. In contrast, the coefficient of variation presents a remarkably different ranking, with low-prevalence patterns (such as J and K) showing significantly higher relative dispersion than high-prevalence patterns (such as B and C). This finding suggests that even low-prevalence patterns may carry crucial information for the analysis of paintings, as visually confirmed in Figure~\ref{fig:2}. It also underscores the importance of caution when comparing raw prevalence values among different paintings and artistic styles, since minor changes can correspond to substantial variations relative to the standard deviation for several patterns.

To account for these distinct scales of variations, we define a standardized measure of prevalence, the $z$-score, as 
\begin{equation*}
z_i = \frac{p_i-\mathbb{E}[{p}_i]}{\mathbb{S}[{p}_i]}\,,
\end{equation*}
where $p_i$ is the probability of finding the $i$-th pattern in a painting, and $\mathbb{E}[{p}_i]$ and $\mathbb{S}[{p}_i]$, represent, respectively, the average and the standard deviation of its prevalence estimated across all images. The value of $z_i$ quantifies how many standard deviation units the prevalence of the $i$-th pattern in a specific painting is above ($z_i>0$) or below ($z_i<0$) its overall incidence across all images. The values of $z_i$ for all 75 patterns are comparable in scale and further allow us to measure how distant the distribution of ordinal patterns of a painting is from the average prevalence. Specifically, we define this distance to the average distribution as the absolute sum of the $z$-score probabilities $s=\sum_{i=1}^{75}|z_i|$. Indeed, we have already used the values of $s$ to color-code the UMAP projection in Figure~\ref{fig:2}, with lighter shades representing small distances and darker shades indicating large distances. Thus, in this representation, the proximity of paintings to the average pattern decreases as they move further from the center of the spiral-like shape. For instance, Candido Portinari's ``Colhedores de Café'' (1935) is near the center of the spiral with $s\approx16$, while Ellsworth Kelly's ``Red Blue Green'' (1963) is positioned at one end of the spiral with $s\approx113$.

As the values of $s$ encapsulate information from all ordinal patterns, they provide an opportunity to globally assess whether the heterogeneous quality of digital reproductions in our dataset introduces any clear bias in our analysis. To examine this aspect, we estimate the association between the distance to the average pattern $s$ and the total number of pixels for each image in our dataset. We also explore the relationship between $s$ and the pixel density for approximately 20\% of the images for which the physical dimensions of the artworks were available on Wikiart. The results indicate no evident association between $s$ and either image size or pixel density (Supplementary Figure~S3), as the correlations between $s$ and these image properties are very weak. These findings suggest that ordinal patterns are not significantly biased by the heterogeneous quality of the images in our dataset. Nevertheless, caution remains essential when examining specific nuances of individual artworks, as illustrated in our prior discussion of Seurat's ``Grande Jatte.''

\begin{figure*}[ht!]
  \centering
  \includegraphics[width=\textwidth,keepaspectratio]{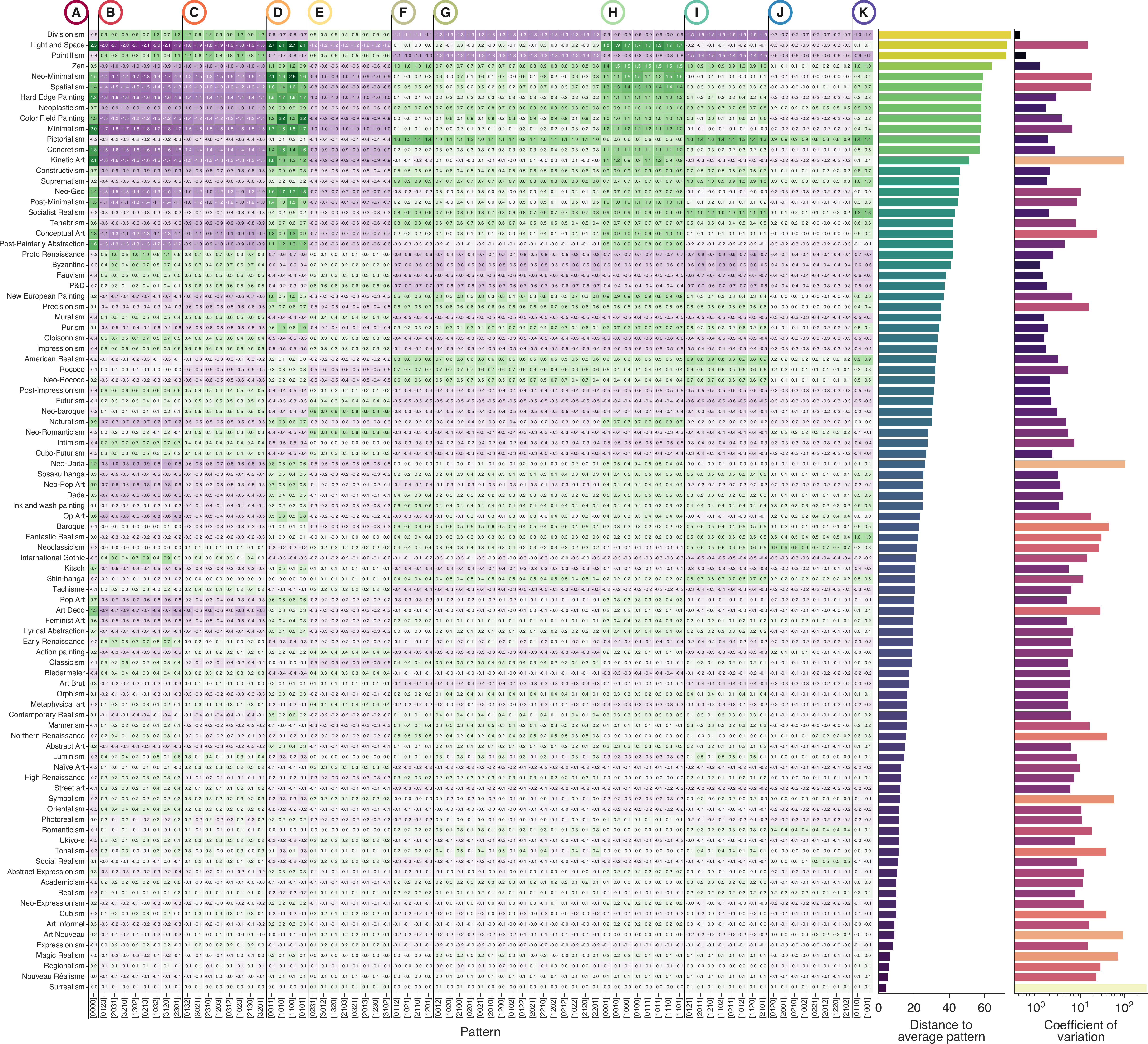}
  \caption{Distinguishing artistic styles using $z$-score probabilities of two-by-two ordinal patterns. The lines in the matrix plot depict the $z$-score probability of each ordinal pattern for 92 artistic styles, each containing at least one hundred artworks (approximately 90\% of data). The $z$-score probability is calculated by subtracting the average probability of an ordinal pattern within a particular style from the overall average probability and dividing the result by the overall standard deviation of that pattern's probability across all images in the dataset. Positive values (represented by shades of green) indicate patterns occurring more frequently than the overall average, whereas negative values (represented by shades of purple) denote patterns occurring less frequently than the overall average, with deviations expressed in standard deviation units. Styles are arranged in descending order based on their distance to the average pattern, determined by the absolute sum of the $z$-score probabilities, which is visualized in a horizontal bar plot on the right of the matrix plot. The right-most panel shows the average of the absolute values of the coefficients of variation of the ordinal probabilities (in $z$-scores units) across all ordinal patterns.
  }
\label{fig:4}
\end{figure*}

Using these $z$-score probabilities, we further investigate whether different artistic styles exhibit distinct incidences of ordinal patterns. We select 92 styles, each represented by at least one hundred paintings, and calculate the average $z_i$ after grouping the images by style. Additionally, we evaluate the average value of $s$ to rank the styles according to their deviation from the average pattern. Figure~\ref{fig:4} depicts these averages in a matrix plot where styles are organized in descending order of $s$, with the latter represented by a horizontal bar plot. Furthermore, to quantify the relative dispersion in the incidence of ordinal patterns within styles, we estimate the coefficients of variation of the $z$-score probabilities of each ordinal pattern, averaging their absolute values across all ordinal patterns. These averaged coefficients of variation for each style are also shown as horizontal bars in Figure~\ref{fig:4}. In the matrix visualization, shades of green indicate patterns exceeding the overall incidence, while shades of pink represent patterns below this incidence. Each row corresponds to the ordinal fingerprint of a style, linking the prevalence of ordinal patterns to the visual characteristics of each style. 

Although a comprehensive analysis of all styles is beyond the scope of our study, several noteworthy global features exist. For example, Divisionism and Pointillism show a lower incidence of types A, D, and H, as well as other patterns featuring identical values along rows or columns (types G and I). In contrast, more discontinuous types, such as B, C, and E, are more prevalent. Despite the limitations of our dataset, these ordinal features seem to reflect the textured visuals and often indistinct edges characteristic of Divisionist and Pointillist compositions. Light and Space, in turn, display almost the opposite behavior, with a higher prevalence of continuous types (such as A, D, and H) and a lower incidence of more discontinuous types (such as types B, C, and E). Notwithstanding the constraints of our data, these ordinal features align with the geometric and well-defined shapes typical of this style. These two styles also differ markedly regarding the degree of dispersion in the prevalence of ordinal patterns, as quantified by their average coefficients of variation, which are significantly higher for the Light and Space style. Most styles, indeed, show higher average coefficients of variation that tend to increase as they approach the overall prevalence of patterns (Supplementary Figure~S4). In addition to the inherent oversimplification involved in representing artworks solely through two-by-two ordinal patterns, this higher variability may also relate to the heterogeneous quality of our dataset and the challenges of obtaining reliable style annotation, which depend on specialized art and art history expertise often lacking in crowdsourcing annotators. This variability further reflects the intrinsic ambiguity of attributing a unique style to artworks, as style evolution tends to be gradual, and paintings may incorporate elements from multiple styles~\citep{elgammal2018shape}. Additionally, certain styles are more narrowly defined by their ``family resemblance,'' whereas others are not. For example, Dada, Neo-Dada, and Surrealism are not generally regarded as cohesive aesthetic doctrines~\citep{gascoyne1970short, hopkins2004dada} and encompass heterogeneous works featuring a wide range of forms and techniques, which may also explain their high coefficients of variation.

\begin{figure*}[ht!]
  \centering
  \includegraphics[width=\textwidth]{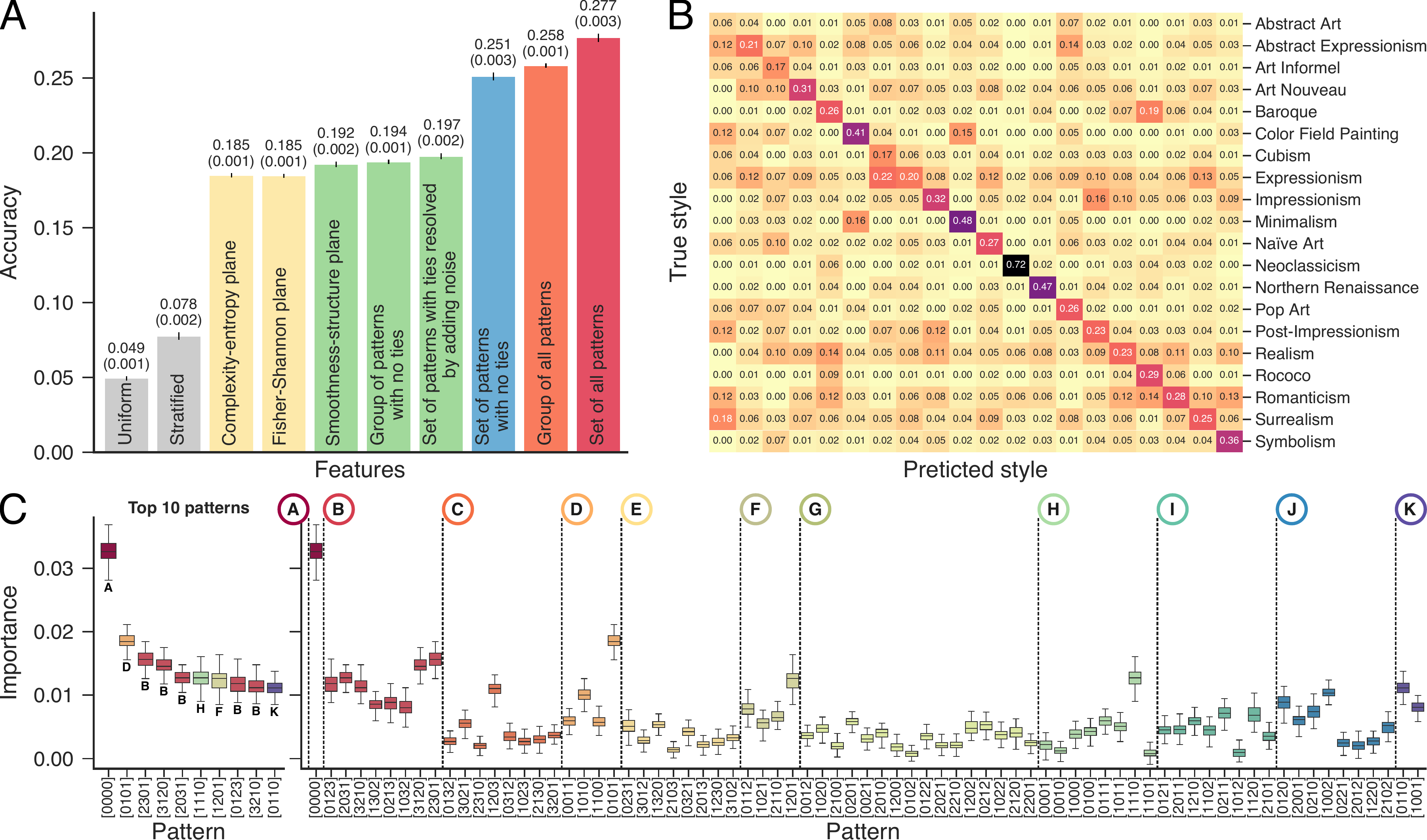}
  \caption{Machine learning artistic styles using two-by-two ordinal patterns. (A) Accuracy of classifiers trained to predict artistic styles from paintings using varied ordinal features. The dataset includes images from 20 styles, each with at least 1,500 artworks, divided into training (80\%) and test sets in a stratified manner. Bars represent the average accuracy across ten independent train-test splits and training procedures, with small error bars showing the standard deviation of accuracy. Numbers above bars indicate average accuracies, with values in parentheses representing standard deviations. Gray bars denote accuracy from dummy classifiers using uniformly random (uniform) and frequency-based (stratified) predictions. Yellow bars depict XGBoost predictions from two pairs of quantifiers derived from the standard ordinal distribution (patterns without ties): the complexity-entropy plane ($H \times C$) and the Fisher-Shannon plane ($H \times F$). Green bars illustrate XGBoost predictions from the smoothness-structure plane ($\kappa\times\tau$), probabilities of pattern groups without ties ($q_1, q_2,$ and $q_3$), and the ordinal distribution with ties resolved by adding small noise to the original images. The blue bar indicates XGBoost predictions from the standard ordinal distribution (patterns without ties). The orange bar represents predictions based on probabilities associated with the eleven pattern types ($q_A, q_B, q_C,\dots,$ and $q_K$). Finally, the red bar shows XGBoost predictions using probabilities related to all 75 two-by-two ordinal patterns. (B) Confusion matrix for the XGBoost model trained with the full set of ordinal probabilities and applied to the test set. (C) Box-plots displaying the permutation importance of each two-by-two ordinal pattern in predicting artistic styles. Permutation importance measures accuracy degradation after randomizing probabilities of an ordinal pattern among all images. The left panel highlights the ten most important patterns, and the right panel displays the importance of each ordinal pattern. Horizontal lines within boxes indicate medians, boxes denote the interquartile range, and whisker bars represent extreme observations.
  }
\label{fig:5}
\end{figure*}

We further investigate the possibility of using the incidence of ordinal patterns to predict the artistic style of paintings through a machine learning approach. In order to have enough training examples, we consider the 100,707 paintings from the 20 most common styles in our dataset, each represented by at least 1,500 artworks. Our task is to determine the style of these images using their distributions of ordinal patterns ($P=\{p_i;~i=1,\dots,75\}$) as predictive features. Additionally, we compare the accuracy of this approach against other ordinal quantifiers that do not consider patterns associated with identical values, namely: the complexity-entropy plane~\citep{sigaki2018history, rosso2007distinguishing, ribeiro2012complexity}, the Fisher-Shannon plane~\citep{olivares2012contrasting}, the smoothness-structure plane~\citep{bandt2023two}, the incidence of the three pattern types defined by Bandt and Wittfeld~\citep{bandt2023two} (types I, II, and III), the probability of all 24 patterns disregarding ties, and the same set of 24 patterns obtained randomly resolving rank ties. These approaches are detailed in the Methods Section~\ref{methods:withouties}. We also aggregate the probabilities of all 75 patterns into their eleven categories and consider the incidence of these pattern types as predictive features for a painting's artistic style. As outlined in the Methods Section~\ref{methods:ml}, we rely on the extreme gradient boosting (XGBoost)~\citep{chen2016xgboost} algorithm to predict the artistic style of paintings using the incidence of ordinal patterns and all other sets of ordinal features. It is important to note that, beyond previously discussed limitations related to style definition, the simplification inherent in reducing paintings to ordinal patterns, and the heterogeneous quality of our dataset, the primary objective of this analysis is not to develop a state-of-the-art style classification system comparable with deep learning approaches~\citep{wu2013painting, zujovic2009classifying, tan2016ceci, mao2017deepart, elgammal2018shape}. Instead, our goal is to assess the informational value of accounting for identical pixel intensities in two-by-two patterns and the effectiveness of categorizing these patterns into eleven distinct types.

Figure~\ref{fig:5}A presents the average accuracy of our predictions and compares these results with those from baseline classifiers that either make uniformly random predictions (uniform) or generate random predictions respecting the style distribution (stratified). All sets of ordinal features achieve accuracy levels significantly higher than those of the baseline classifiers, indicating that ordinal patterns carry stylistic information about paintings, as also discussed by Sigaki \textit{et al.}~\citep{sigaki2018history}. Notably, using all 75 ordinal patterns as predictive features results in a significantly enhanced accuracy compared to all other sets of ordinal features. This improvement is particularly pronounced when compared with the complexity-entropy, Fisher-Shannon, and smoothness-structure planes, as well as the incidence of types I, II, and III, and the set of all 24 ordinal patterns with rank ties resolved randomly. These former approaches all approximate an accuracy of $0.19$, whereas using all two-by-two ordinal patterns achieves $0.28$. Furthermore, the smoothness-structure plane, the incidence of types I, II, and III, and the set of all 24 ordinal patterns with rank ties resolved randomly show only marginally higher accuracy than the complexity-entropy and Fisher-Shannon planes. This result demonstrates that permutation entropy combined with statistical complexity or the Fisher information measure effectively compresses information from ordinal distributions, despite the meticulous theoretical underpinnings of the smoothness-structure plane and patterns types I, II, and III~\citep{bandt2023two}. Additionally, randomly resolving rank ties yields lower accuracy than the set of 24 ordinal patterns without ties, which achieves an accuracy of $0.25$. While identical values are not explicitly addressed in this latter approach, their occurrences are deterministically handled based on the positions of equal values, thereby retaining additional ordering information rather than randomly resolving ties. Interestingly, although comprising less than half the dimensionality of the set of 24 ordinal patterns without ties, the incidence of the eleven pattern types yields the second-highest accuracy. This finding indicates that these pattern groups effectively compress the information encoded in all 75 two-by-two ordinal patterns. However, despite the symmetries within each pattern type, the lower performance of classifiers trained with these grouped patterns suggests that each ordinal pattern provides unique and critical information for identifying the style of a painting.

We also calculate the confusion matrix for the classifier trained with all two-by-two ordinal patterns, as shown in Figure~\ref{fig:5}B. As expected, the overall accuracy of approximately 28\% achieved with our ordinal features is modest compared to more sophisticated machine learning approaches, such as convolutional neural networks, which attain accuracies above 70\% in similar tasks~\citep{zujovic2009classifying, tan2016ceci, mao2017deepart, elgammal2018shape}. Nonetheless, our approach demonstrates that the 75 ordinal patterns contain critical information for classifying styles that are not present in the standard distribution of ordinal patterns nor in its usual entropy measures. Moreover, the confusion matrix exhibits a well-defined diagonal structure, indicating that the most frequent predictions across all styles align with the actual styles of images. For instance, the trained algorithm correctly identified Neoclassicism, Minimalism, and Northern Renaissance artworks in 72\%, 48\%, and 47\% of the predictions, respectively. However, our model struggles with Abstract Art, correctly identifying this style in only 6\% of predictions. The low performance with Abstract Art highlights the difficulties in assigning a single style to works encompassing a diverse range of techniques, which often lack distinctive, consistent features and can vary widely among artists and even within a single artist's oeuvre. Similar observations were made by Karjus \textit{et al.}~\citep{karjus2023compression}, who used compression ensembles to classify styles and noted that certain styles are more frequently misclassified, suggesting that the structure of these errors may hold interpretive value. Additionally, our algorithm tends to misclassify more styles that exhibit high coefficients of variation in the prevalence of ordinal patterns and low distance to the average pattern (such as Cubism and Art Informel), while styles with low coefficients of variation and greater distance from the average pattern (such as Minimalism and Color Field Painting) are more frequently correctly identified.

To better understand the role of each ordinal pattern in predicting styles, we evaluate the importance of all patterns by measuring their contribution to the overall accuracy of the model (see Methods Section~\ref{methods:ml}). Figure~\ref{fig:5}C shows the importance measure for each ordinal pattern grouped into their eleven categories, as well as the top 10 patterns that most significantly contribute to the overall accuracy. There is no evident association between pattern types or overall prevalence of patterns (see also Figure~\ref{fig:1}) and their importance in classifying artistic styles. Notably, the ranking of the most important patterns includes the pattern $[0110]$, which belongs to the least prevalent type (K). Another remarkable finding is the disproportional importance of the type A pattern $[0000]$, which alone accounts for 12\% of the overall accuracy. It is equally intriguing to note that five type B patterns are the only patterns without identical values in the top 10 importance rank, the remaining five patterns are from types with identical values. Combined with the low classification performance in models trained with ordinal patterns in which identical values are randomly resolved, these findings further corroborate the significance of adequately addressing identical values when investigating ordinal patterns in images.

\begin{figure*}[ht!]
  \centering
  \includegraphics[width=1\textwidth]{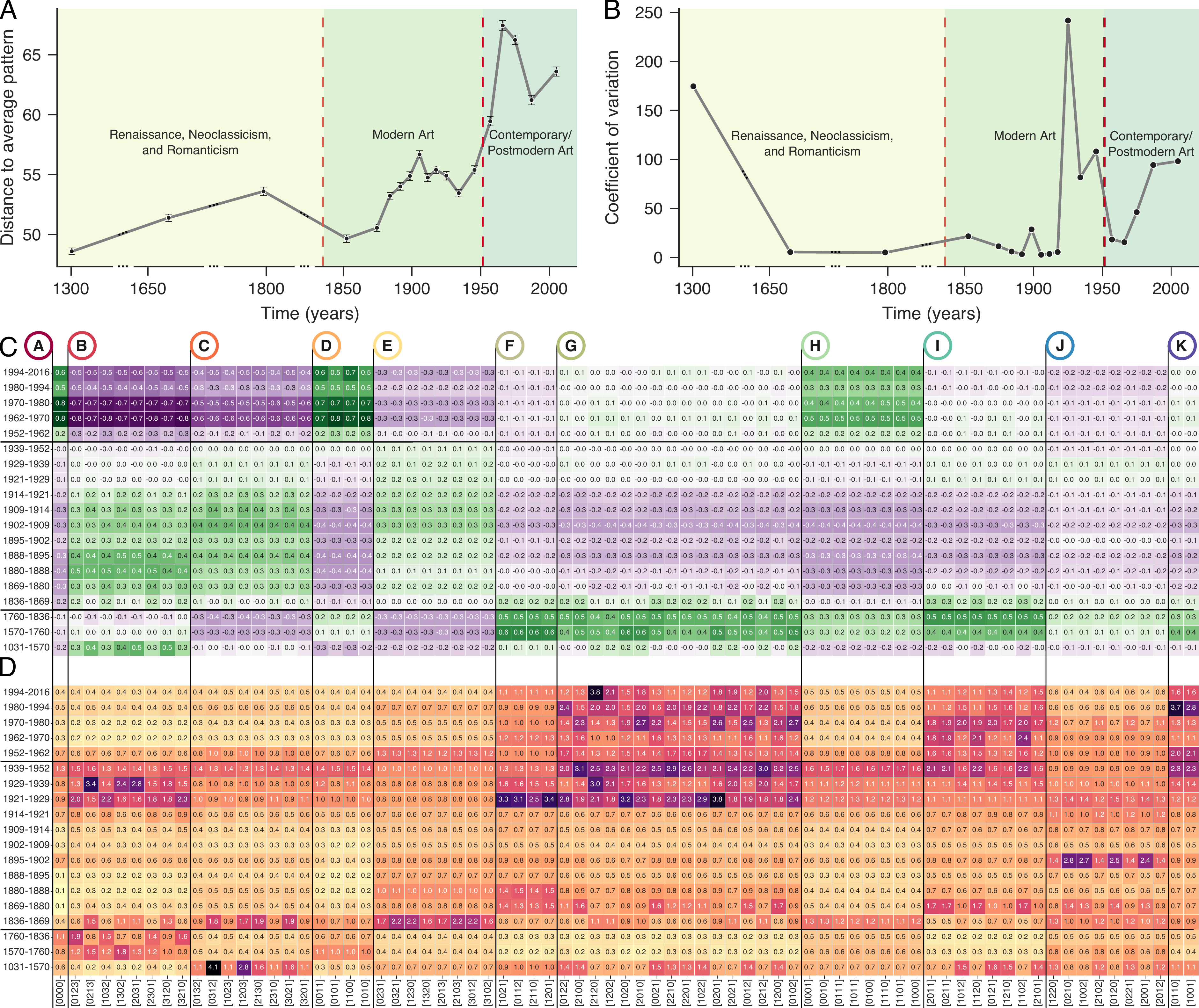}
  \caption{Evolution of paintings analyzed by the distance to the average ordinal pattern. (A) Circles indicate the mean distances to the average ordinal pattern, calculated by grouping artworks by date; error bars represent the standard error of the mean. (B) Absolute sum of the coefficients of variation for each ordinal pattern, calculated after grouping artworks by date. The temporal intervals used to calculate the averages and coefficients of variation contain nearly equivalent numbers of images and are not uniformly distributed over time. Both panels (A) and (B) feature broken temporal axes, indicated by ellipsis points, to represent gaps in time and enhance the visualization of the temporal evolution. Additionally, vertical lines, as well as colored backgrounds, delineate major art historical periods: Medieval Art, the Renaissance, Neoclassicism, and Romanticism, which developed until the 1850s; Modern Art, initiated with the advent of Impressionism in the 1870s and characterized by the rise of avant-garde styles such as Cubism, Expressionism, and Surrealism in the early 20th century; and Contemporary/Postmodern Art, typically associated with the emergence of Pop Art in the 1960s. (C) Matrix plot displaying the $z$-score probability of each ordinal pattern calculated within each temporal interval. This quantity is calculated by subtracting the average probability of an ordinal pattern in a given period from its overall average and dividing the result by the overall standard deviation of that pattern's probability across all images in the dataset. Shades of green indicate patterns occurring more frequently than the overall average, while shades of purple represent patterns occurring less frequently. (D) Coefficients of variation of the ordinal probabilities (in $z$-scores units) of each ordinal pattern calculated within each temporal interval. The matrix plot shows the base-10 logarithm of the absolute values of the coefficients of variation, with darker shades representing larger values. Horizontal lines in panels (C) and (D) delineate the same major art historical periods as in panels (A) and (B).
  }
\label{fig:6}
\end{figure*}

In the final aspect of our investigation, we explore the potential of quantifying the temporal evolution of art paintings by examining changes in the prevalence of ordinal patterns. This approach builds on the findings of Sigaki \textit{et al.}~\citep{sigaki2018history}, who quantified the evolution of art paintings using the complexity-entropy plane. However, beyond previously discussed challenges related to the quality and resolution of digital reproductions available in Wikiart, it is crucial to recognize that tracking these patterns over time with this dataset is inherently limited due to its underlying biases. As noted in Refs.~\citep{lee2020dissecting, karjus2023compression, elgammal2018shape}, the Wikiart collection largely consists of Western European artworks, reflecting the historical canon and thus providing a skewed representation of the global evolution of art. Moreover, it displays notable biases towards male artists~\citep{lee2020dissecting}, as well as towards 20th-century and post-World War II works, with gaps in coverage for the 18th century and early 19th century~\citep{karjus2023compression, elgammal2018shape}. These biases restrict the diversity of perspectives captured and narrow the temporal scope of our analysis. Consequently, conclusions about art's evolution will only become more robust through the development of balanced, high-quality datasets that are yet to be realized.

With these limitations in mind, we group paintings in our dataset by composition date and calculate the average value of $s$ within time intervals encompassing a comparable number of artworks. Figure~\ref{fig:6}A depicts the evolution of the average value of $s$, with different background colors delineating three major art historical periods: Medieval Art, the Renaissance, Neoclassicism, and Romanticism, which developed until the 1850s; Modern Art, initiated with the onset of Impressionism in the 1870s and marked by the emergence of avant-garde styles such as Cubism, Expressionism, and Surrealism in the early 20th century; and Contemporary/Postmodern Art, typically associated with the rise of Pop Art in the 1960s~\citep{danto1997after}. Each period exhibits distinctive typical distances to the average ordinal pattern, corroborating the observation that artistic styles also display different values of $s$. Notably, the evolution of art paintings shows a general trend of increasing divergence from the average ordinal pattern. This divergence accelerated during the transition from Modernism to Postmodernism, a pattern also observed in the complexity-entropy plane of Sigaki \textit{et al.}~\citep{sigaki2018history}. In addition to the average value of $s$, we calculate the absolute sum of the coefficients of variation for each ordinal pattern within the same time intervals. Figure~\ref{fig:6}B indicates that the evolution of art, as depicted by changes in $s$, is neither smooth nor homogeneous, with substantial variations in the prevalence of ordinal patterns. This variability is particularly pronounced in the earliest period in our dataset and after the 1930s. While the high variability in the earliest period partially reflects the broad range of epochs it encompasses -- aggregating artworks from diverse historical contexts -- the pronounced variability during the transition from Modernism to Postmodernism aligns with interpretations of this shift as marked by stylistic disruptions rather than a gradual evolution, leading to a fragmented, diverse artistic landscape in which multiple styles and approaches coexist~\citep{foster2016art, mesoudi2018cumulative, karjus2023compression}. We further examine individual changes in the prevalence of ordinal patterns by calculating their $z$-score probabilities relative to the average prevalence across all images, as shown in Figure~\ref{fig:6}C, and the coefficients of variation of each ordinal pattern, as presented in Figure~\ref{fig:6}D (on logarithmic scale). These results corroborate the trends observed when aggregating all patterns to estimate the values of $s$ but also highlight some interesting shifts in the prevalence of ordinal patterns, primarily marked by an increase in the incidence of patterns A, D, and H and a decrease in type B in more recent works (despite early artworks also exhibiting positive $z$-scores for types D and H). However, once again, the often high variability in the prevalence of ordinal patterns underscores that these shifts do not represent a smooth development.

\section{Discussion and Conclusions}

We have conducted a comprehensive analysis of artworks using ordinal patterns extracted from two-by-two-pixel partitions of images. Initially, we have categorized the 75 possible ordinal patterns into eleven categories based on symmetry considerations and smoothness metrics. These categories were hierarchically organized based on the number of unique symbols and the smoothness of their patterns. Subsequently, each of the nearly 140 thousand images in our dataset was represented by a discrete probability distribution, reflecting the relative frequency of each ordinal pattern. By integrating expertise from the art domain, our method demonstrated the potential of using the distribution of ordinal patterns to facilitate quantitative analyses of various visual aspects of artworks, ranging from the dominance of vertical and horizontal structures to the prevalence of sharp edges and well-defined elements.

Our analysis identified a universal pattern in the occurrence of ordinal patterns across painting images, which led to our classification into eleven distinct pattern types. The average prevalence of these patterns within each group remains approximately constant but monotonically decreases as the roughness of the patterns increases among types sharing the same number of symbols. We have also observed that the incidence rank of pattern types reflects a nontrivial interplay between their smoothness and the decreased likelihood of finding pixels with identical intensities in two-by-two image partitions. Although a clear hierarchy exists in the occurrence of ordinal patterns in paintings, we have verified that patterns with lower prevalence exhibit significant variability compared to their mean probability of occurrence. This finding shows that tiny changes in pattern incidence can lead to substantial relative variations, prompting us to establish a standardized measure for the prevalence of ordinal patterns. Moreover, when summed absolutely, this measure provides a distance measure relative to the average pattern incidence.

We have used this standardized metric to define and rank the ordinal fingerprint of artistic styles based on their proximity to the overall incidence of ordinal patterns. The average incidence of ordinal patterns captures low-level features that appear to correlate with certain aesthetic attributes of styles. For instance, Divisionist and Pointillist compositions, which often depict textured visuals with indistinct edges, show a lower incidence of patterns with identical pixel intensities, whereas Light and Space artworks, characterized by geometric and well-defined shapes, display a higher occurrence of these patterns. Styles also exhibit notable variability in the prevalence of ordinal patterns, highlighting not only the limitations in representing artworks solely through two-by-two ordinal patterns but also the heterogeneity of our dataset, challenges in achieving reliable style annotation, and the inherent ambiguity in assigning a single style to artworks. Despite these complexities, our analysis demonstrated that the prevalence of ordinal patterns is predictive of artistic styles in paintings, and that by adequately accounting for patterns of equal pixel intensity, we have substantially enhanced the performance of style classification. Notably, patterns characterized by identical pixel intensities are among the most important features for style classification, with the type A pattern alone contributing to 12\% of the overall accuracy. {Although our overall accuracy of approximately 28\% is modest compared to deep learning-based approaches~\citep{wu2013painting, zujovic2009classifying, tan2016ceci, mao2017deepart, elgammal2018shape}, it underscores that simple, interpretable, low-level ordinal patterns provide meaningful insights into the features of artworks, particularly when patterns with identical pixel intensities are adequately considered.}

Our investigation also explored the temporal changes of art paintings in our dataset, examining shifts in the prevalence of ordinal patterns and their deviation from the average pattern. The results revealed a general trend wherein art paintings increasingly diverge from the average pattern, punctuated by brief intervals in which this trend is reversed. This evolution is neither smooth nor homogeneous; instead, it is marked by substantial variations in the prevalence of ordinal patterns, particularly in artworks produced after the 1930s and during the transition from Modernism to Postmodernism. These findings suggest that the evolution of art does not follow a unilinear path. Rather, shifts in the prevalence and variability of ordinal pattern prevalence, particularly during transitional periods, appear to align with stylistic disruptions, resulting in a fragmented and diverse artistic landscape where multiple styles and approaches coexist~\citep{karjus2023compression, foster2016art, mesoudi2018cumulative}.

Our work is not without limitations, most of which have been previously acknowledged; however, it is useful to summarize them here. First, our dataset, sourced from Wikiart, consists of a large yet heterogeneously compiled collection of digital reproductions, primarily reflecting Western paintings, which limits its generalizability and introduces variability in image quality and metadata accuracy. The diverse resolutions of artworks, especially larger pieces, further impact our analysis, as high-resolution images capture fine details that can influence the prevalence of certain ordinal patterns. Although we did not observe any clear bias in the association between image size, pixel density, and distance from the average ordinal pattern, caution remains essential, particularly in analyses of nuanced features in individual artworks. Additionally, style classification is inherently complex, with ambiguities in style attribution, variability in stylistic consistency across certain artistic movements, and limited representation within the dataset of non-Western and diverse gender perspectives. The temporal evolution of art is also challenging to track accurately due to dataset biases, as Wikiart predominantly features 20th-century Western works, which may distort broader historical interpretations. Lastly, we must acknowledge that the simplification involved in reducing the rich complexity of paintings to ordinal patterns at a single spatial scale inherently overlooks multiscale details essential to the artworks' aesthetic and structural nuances. Future research, ideally utilizing high-resolution reproductions, could extend our use of ordinal patterns across multiple scales~\citep{zunino2016discriminating}, similar to wavelet transforms, which have proven valuable in art investigations, including dating, artistic identification, authentication, image separation, cradle removal, and brushstroke analysis~\citep{jafarpour2009stylistic, johnson2008image, polatkan2009detection, yin2014digital, deligiannis2016multi}.

Despite these limitations, our research demonstrates that, even amidst the visual complexity of art paintings, simple ordinal patterns emerging from two-by-two-pixel partitions encode critical information capable of suggesting artistic styles, fostering in-depth analysis of their aesthetic features, and even revealing the dynamical behavior of art.

\section{Methods}

\subsection{Data}\label{methods:data} 

The dataset comprises 137,364 digitized paintings from the WikiArt visual arts encyclopedia, as originally introduced by Sigaki \textit{et al.}~\citep{sigaki2018history}. It spans nearly a millennium (1031–2016) and includes works by 2,391 artists representing 154 artistic styles. For the temporal analysis, 33,724 artworks lacking composition dates are excluded (Figure~\ref{fig:6}). Images are stored in JPEG format with 24-bit RGB encoding. Color layers are converted to grayscale for analysis using the standard luminance transformation~\citep{scikit-image}. See the Supplementary Material (Section~1A) for more details.

\subsection{Ordinal patterns without ties}\label{methods:withouties} 
We follow the seminal approach of Bandt and Pompe~\citep{bandt2002permutation} to encode images using ordinal patterns. Images are sampled using sliding partitions; each partition is flattened, and its elements are replaced by their ranks to form an ordinal pattern. However, in contrast to our approach, identical values within a partition are handled by preserving their order of occurrence or adding a small noise term to resolve rank ties without altering other ordering relationships (alternative approaches have been proposed~\citep{bian2012modified}). This procedure yields an ordinal probability distribution $P=\{p_i;~i=1,\dots,n\}$, where each $p_i$ corresponds to the probability of finding one of the $n=(d_x d_y)!=24$ possible ordinal patterns. Thus, our distinction of patterns with identical values introduces 51 additional ordinal patterns, raising the total to 75 possible patterns. 

Using the standard ordinal probability distribution, we calculate the normalized Shannon entropy $H$~\citep{bandt2002permutation}, the statistical complexity~\citep{lopez1995statistical, lamberti2004intensive, martin2006generalized, rosso2007distinguishing, zunino2016discriminating} $C$, and Fisher information measure $F$. Furthermore, following Bandt and Wittfeld~\citep{bandt2023two}, the $24$ ordinal patterns are classified into three types (I, II, III) based on continuity, to then calculate the smoothness $\tau=q_1-1/3$ and the branching structure $\kappa=q_2-q_3$, where $q_1$, $q_2$, and $q_3$ are the relative frequencies of types I, II, and III, respectively. The combination of $H$ and $C$ defines the complexity–entropy plane, the values of $H$ and $F$ yield the Fisher–Shannon plane, and the values of $\tau$ and $\kappa$ determine the smoothness–structure plane. We implement all these ordinal methods using the Python module \textit{ordpy}~\citep{pessa2021ordpy}. See the Supplementary Material (Sections~1B-1E) for further details.

\subsection{Machine learning artistic styles}\label{methods:ml}
We apply the extreme gradient boosting (XGBoost)~\citep{chen2016xgboost} algorithm to classify the 20 most common styles in our dataset, with each style represented by at least 1,500 artworks. We partition the data in a stratified manner -- 80\% for training and 20\% for testing -- and repeat this procedure across ten independent realizations to compute mean and standard deviation of accuracy. We investigate eight sets of ordinal features. Six sets use the Bandt and Pompe encoding~\citep{bandt2002permutation}, comprising the complexity–entropy, Fisher–Shannon, and smoothness–structure planes; the probabilities of the three pattern types; the probabilities of the 24 possible patterns without ties; and the corresponding probabilities obtained by resolving rank ties through the addition of small noise. The remaining two feature sets, original to our work, explicitly consider the occurrence of identical values within the image partitions: the probability associated with each of the eleven pattern types ($q_A, q_B, q_C,\dots,$ and $q_K$) and the probability of each of the 75 ordinal patterns. Finally, we assess the contribution of each of the 75 ordinal patterns using permutation feature importance~\citep{breiman2001random, fisher2019all, molnar2020interpretable} as implemented in scikit-learn~\citep{pedregosa2011scikitlearn}. See the Supplementary Material (Section~1F) for further details.

\subsection{Low-Dimensional Projection of the Ordinal Pattern Probabilities}\label{methods:UMAP}
We use uniform manifold approximation and projection (UMAP)~\citep{mcinnes2018umap} to reduce 75-dimensional ordinal pattern frequency vectors to two dimensions. UMAP constructs a weighted graph (fuzzy simplicial complex) from a dissimilarity matrix and projects the data via a force-directed layout algorithm. Owing to inherent stochasticity, projections vary slightly between runs, rendering UMAP most useful for comparative visualization. We implement this approach using the Python package \textit{umap}~\citep{mcinnes2018umapsoftware}. See the Supplementary Material (Section~1G) for further details.

\section{Funding}
This work is supported in part by funds from the Coordena\c{c}\~ao de Aperfei\c{c}oamento de Pessoal de N\'ivel Superior (CAPES), the Conselho Nacional de Desenvolvimento Cient\'ifico e Tecnol\'ogico (CNPq -- Grant 303533/2021-8), the Consejo Nacional de Investigaciones Científicas y Técnicas (CONICET, Argentina), and the Slovenian Research Agency (Grant Nos. P1-0403 and N1-0232). 

\section{Author contributions statement}
All the authors designed and performed the research and wrote and reviewed the manuscript.

\section{Competing interest statement}
The authors declare no competing interest.

\section{Data availability}
The data that support the findings of this study are available as a supplementary file (File S1).

\bibliography{references}



\section*{Supplementary material}

# Two-by-two ordinal patterns in art paintings

## (Supplementary Material, PNAS Nexus, 2025)

Mateus M. Tarozo,[1] Arthur A. B. Pessa,[1] Luciano Zunino,[2, 3]
Osvaldo A. Rosso,[4] Matjaž Perc,[5, 6, 7, 8, 9] and Haroldo V. Ribeiro[1, *]

[1]*Departamento de Física, Universidade Estadual de Maringá – Maringá, PR 87020-900, Brazil*
[2]*Centro de Investigaciones Ópticas (CONICET La Plata - CIC - UNLP), 1897 Gonnet, La Plata, Argentina*
[3]*Departamento de Ciencias Básicas, Facultad de Ingeniería,
Universidad Nacional de La Plata (UNLP), 1900 La Plata, Argentina*
[4]*Instituto de Física, Universidade Federal de Alagoas – Maceió 57072-900, Brazil*
[5]*Faculty of Natural Sciences and Mathematics, University of Maribor, Koroška cesta 160, 2000 Maribor, Slovenia*
[6]*Community Healthcare Center Dr. Adolf Drolc Maribor, Ulica talcev 9, 2000 Maribor, Slovenia*
[7]*Department of Physics, Kyung Hee University, 26 Kyungheedae-ro,
Dongdaemun-gu, Seoul 02447, Republic of Korea*
[8]*Complexity Science Hub, Metternichgasse 8, 1030 Vienna, Austria*
[9]*University College, Korea University, 145 Anam-ro, Seongbuk-gu, Seoul 02841, Republic of Korea*

## SUPPORTING INFORMATION TEXT

### 1. DETAILED METHODS

#### A. Data

The dataset of paintings used in our investigation is the same as that introduced by Sigaki *et al.* [1] and comprises 137,364 digitized images, along with metadata related to each artwork obtained from the visual arts encyclopedia WikiArt [2]. These images cover works from 2,391 artists and 154 artistic styles, spanning a timeline of nearly a millennium (from 1031 to 2016). A total of 33,724 artworks with missing composition dates were excluded from the analysis of the temporal evolution (Figure 6). All files are stored in JPEG format with 24 bits per pixel, encompassing 8 bits for each of the red, green, and blue colors in the RGB color space. For our analyses, the three color layers of each image were converted to grayscale using the standard luminance transformation [3], resulting in a simple matrix with entries corresponding to pixel intensities calculated as $0.2125R + 0.7154G + 0.0721B$, where $R$, $G$ and $B$ represent the intensities of the red, green, and blue layers.

#### B. Ordinal patterns without ties

We revisit the seminal ordinal encoding proposed by Bandt and Pompe [4] in the context of image data [5, 6] by once again considering the matrix

$$A = \begin{bmatrix} 4 & 2 & 9 \\ 5 & 3 & 1 \\ 2 & 6 & 3 \end{bmatrix}, \qquad (1)$$

representing a hypothetical image of width $n_x = 3$ and height $n_y = 3$, where elements correspond to the pixel intensities. As in the maintext, we sample this image using a sliding matrix of dimensions $d_x = 2$ per $d_y = 2$ (the so-called embedding dimensions) that traverses $A$ in unitary steps along both the horizontal and vertical axes, resulting in four partitions:

$$A_0 = \begin{bmatrix} 4 & 2 \\ 5 & 3 \end{bmatrix}, \quad A_1 = \begin{bmatrix} 2 & 9 \\ 3 & 1 \end{bmatrix}, \quad A_2 = \begin{bmatrix} 5 & 3 \\ 2 & 6 \end{bmatrix}, \text{ and } \quad A_3 = \begin{bmatrix} 3 & 1 \\ 6 & 3 \end{bmatrix}.$$

---

* hvr@dfi.uem.br

We subsequently flatten these submatrices into sequences $(a_0, a_1, a_2, a_3)$ and replace the elements with their respective ranks within each submatrix, thereby defining an ordinal pattern reflective of the partition's motif. Thus, the partition $A_0$ yields the ordinal pattern [2031] because $a_0 = 4$ is the second-largest element (rank 2), $a_1 = 2$ is the smallest element (rank 0), $a_2 = 5$ is the largest element (rank 3), and $a_3 = 3$ is third-largest element (rank 1). Similarly, the patterns for $A_1$ and $A_2$ are [1320] and [2103], respectively. Differently from our approach, the standard procedure for handling identical values within a partition is to assign ranks based on their positions. Hence, $A_3$ is described by the ordinal pattern [1032] because the first number 3 is at the $a_0$ position, whereas the second occurrence is at the $a_3$ position (we note that $A_3$ corresponds to [1021] when properly accounting for the equality). Another option for handling these identical values is to add a small noise term to resolve rank ties without altering other ordering relationships – a scheme suggested by Bandt and Pompe [4] but less commonly used in the literature. In this case, $A_3$ may correspond either to the ordinal pattern [1032] or to [2031], depending on whether the added noise disrupts the equality between $a_0$ and $a_3$ either $a_0 < a_3$ or $a_0 > a_3$.

Furthermore, the original formulation of Bandt and Pompe [4] does not use ranking to define the ordinal patterns. Instead, it evaluates permutations $\pi = (r_0, r_1, r_2, r_3)$ of index numbers $(0, 1, 2, 3)$, which organize the elements $(a_0, a_1, a_2, a_3)$ in ascending order, defined by $a_{r_0} \leq a_{r_1} \leq a_{r_2} \leq a_{r_3}$. In instances of equal values, the order of occurrence among partition elements is preserved. For instance, using this method, $A_0$ is described by $\pi = (1, 3, 0, 2)$ as $a_1 < a_3 < a_0 < a_2$ $(2 < 3 < 4 < 5)$, while $A_3$ is characterized by $\pi = (1, 0, 3, 2)$ because $a_1 < a_0 \leq a_3 < a_2$ $(1 < 3 \leq 3 < 6)$. Naturally, when no identical values exist, the rank-based and the permutation-based approaches yield the same number of possible patterns, and there is a one-to-one correspondence between the permutations $\pi$ and the ordinal patterns derived from ranking the partition values. Indeed, we can find these ordinal patterns by simply finding the permutations $\tilde{\pi} = (\tilde{r}_0, \tilde{r}_1, \tilde{r}_2, \tilde{r}_3)$ that sort $\pi = (\tilde{a}_0, \tilde{a}_1, \tilde{a}_2, \tilde{a}_3)$ in ascending order; for example, $\pi = (1, 3, 0, 2)$ correspond to [2031] because $\tilde{a}_2 < \tilde{a}_0 < \tilde{a}_3 < \tilde{a}_1$.

Both approaches thus allow us to represent images $A$ by the same ordinal probability distribution $P = \{p_i;\ i = 1, \dots, n\}$, where only the order of the elements in $P$ changes between the two encoding procedures. Each $p_i$ corresponds to the probability of finding each of the $n = (d_x d_y)! = 24$ possible ordinal patterns across the $(n_x - d_x + 1)(n_y - d_y + 1) = 4$ partitions of $A$. Notably, the distinction of patterns with identical values introduces 51 additional ordinal patterns, raising the total to 75 possible patterns. We also note that Bian *et al.* [7] proposed a procedure to address equal values in the context of heartbeat time series analysis. Their approach evaluates permutations as in the original formulation but maps equal values into the same symbol corresponding to the smallest index number of the tied values. However, the number of possible patterns derived from Bian *et al.*'s approach is smaller than in our rank-based methodology, as certain rank patterns map to identical permutation patterns. For example, [1010] and [1001] both correspond to $\pi = (1, 1, 0, 0)$, while [0101] and [0110] map to $\pi = (0, 0, 1, 1)$ in Bian *et al.*'s scheme.

## C. Complexity-entropy plane

Using the ordinal distribution $P = \{p_i;\ i = 1, \dots, n\}$, we can calculate several quantifiers. The most celebrated is the normalized Shannon entropy

$$H(P) = \frac{1}{\ln(n)} \sum_{i=0}^{n} p_i \ln(1/p_i), \tag{2}$$

that defines the permutation entropy [4]. The $H$ values quantify the degree of disorder in the ordering of pixel intensities, where $H \approx 1$ corresponds to a random pattern and $H \approx 0$ indicates that pixels consistently assume the same order. It is also common to combine the values of $H$ with other quantifiers, such as the statistical complexity $C$ [8–10] or Fisher information measure $F$ [11, 12].

Statistical complexity is defined as

$$C(P) = \frac{D(P, U) H(P)}{D^*}, \tag{3}$$

where $D(P, U)$ represents the Jensen-Shannon divergence between the ordinal distribution $P$ and the uniform distribution $U = \{u_i = 1/n;\ i = 1, \dots, n\}$, calculated via

$$D(P, U) = S\left(\frac{P + U}{2}\right) - \frac{S(P)}{2} - \frac{S(U)}{2}, \tag{4}$$

with $\frac{P+U}{2} = \left\{\frac{p_i + 1/n}{2},\ i = 1, \dots, n\right\}$ and

$$D^* = -\frac{1}{2}\left[\frac{n+1}{n}\ln(n+1) + \ln(n) - 2\ln(2n)\right] \tag{5}$$

representing a normalization constant corresponding to the maximum value of $D(P,U)$. The value of $C(P)$ is zero in both extremes of order ($P = \{p_i = \delta_{i^*,i};\ i = 1,\ldots,n\}$ for some $i^*$ between 1 and $n$) and disorder ($P = \{p_i = 1/n;\ i = 1,\ldots,n\}$); however, $C(P)$ is not a trivial function of $H(P)$. For a given $H(P)$, $C(P)$ can range between minimum and maximum values [6, 10, 13], carrying information not present in the permutation entropy. The combined use of $C(P)$ versus $H(P)$ yields a diagram often referred to as the complexity-entropy plane [5, 6, 8, 13].

### D. Fisher-Shannon plane

The Fisher information measure is given by [14]

$$F(P) = F^* \sum_{i=1}^{n-1} \left(\sqrt{p_{i+1}} - \sqrt{p_i}\right)^2, \tag{6}$$

where

$$F^* = \begin{cases} 1 & \text{if } p_{i^*} = 1 \text{ for } i^* = 1 \text{ or } i^* = n \\ 1/2 & \text{otherwise} \end{cases} \tag{7}$$

is a normalization constant ensuring that $0 < F < 1$. In this calculation, the elements $p_i$ of the ordinal probability distribution are sorted in lexicographic order as determined by the Lehmer code [15, 16]. The value of $F$ is a local measure highly sensitive to changes among the elements of the ordinal distribution, akin to the gradient of a probability distribution in the continuous case. The Fisher information measure exhibits behavior antithetical to permutation entropy, approaching one in highly ordered states ($F \approx 1$) and zero in disordered states ($F \approx 0$). Moreover, similarly to statistical complexity, the value of $F$ is not related to $H$, which motivates the joint use of both quantifiers in a diagram commonly referred to as the Fisher-Shannon plane [14].

### E. Smoothness-structure plane

Beyond the aforementioned and other entropy-related quantifiers, Bandt and Wittfeld [17] have recently introduced a novel method for the ordinal analysis of images, resulting in the definition of two metrics named smoothness and branching structure. This methodology, which was the main inspiration for our investigation of paintings, proposed analyzing each of the 24 possible ordinal patterns emerging from two-by-two partitions of images. Bandt and Wittfeld categorized these patterns into three groups, each containing eight patterns, referred to as type I, type II, and type III, and distinguished by symmetry and roughness characteristics. These patterns respectively comprise our types B, C, and E; the three categories consisting of patterns containing four distinct symbols, as illustrated in Figure 1.

In our notation, the type I group comprises the primary patterns [0123] and [0213], along with their rotated versions by quarter ([2031] and [1032]), half ([3210] and [3120]), and three-quarter ([1302] and [2301]) turns. This group contains the most continuous ordinal patterns, which emerge when corresponding pixel intensities lie in a three-dimensional plane (simultaneously increasing or decreasing along columns and rows). Type II are formed by the primary patterns [0132] and [0312], and their rotated versions by quarter ([3021] and [1023]), half ([2310] and [2130]), and three-quarter ([1203] and [3201]) turns. These patterns are less continuous than those of Type I and arise when pixel intensities within the partitions simultaneously increase or decrease along one direction but show distinct trends along the other (for instance, in [0132], the first row increases from left to right while the second decreases, with values along both columns increasing from top to bottom). Type III, the most discontinuous patterns, comprises the primary patterns [0231] and [0321], and their rotated versions by quarter ([3012] and [2013]), half ([1320] and [1230]), and three-quarter ([2103] and [3102]) turns. In these patterns, pixel intensities within the partitions exhibit simultaneously distinct trends along rows and columns (for instance, in [0231], the first row increases from left to right while the second decreases, and the first column increases from top to bottom while the second decreases).

Using this classification of ordinal patterns, Bandt and Wittfeld calculate the relative frequency of each pattern type, defining the smoothness

$$\tau = q_1 - 1/3 \tag{8}$$

and the branching structure as

$$\kappa = q_2 - q_3 \tag{9}$$

where $q_1$, $q_2$, and $q_3$ denote the relative frequencies of type I, II, and III ordinal patterns, respectively. The values of $\tau$ quantify the smoothness/roughness of the pixel intensities within the partitions, reaching a maximum value of 2/3 when intensities consistently increase or decrease along both directions (only type I patterns emerge) and a minimum of $-1/3$ when intensities always display distinct trends along both directions (only type III patterns emerge). This measure is analogous to the persistence parameter defined for ordinal patterns in time series [18]. In its turn, the values of $\kappa$ range from $-1$ (only type III patterns emerge) to 1 (only type II patterns emerge), quantifying the presence of branching and spiral-like patterns in pixel intensities. Bandt and Wittfeld also introduced the joint use of $\tau$ and $\kappa$ to characterize images, defining a diagram that we termed the smoothness-structure plane.

The Python module *ordpy* [19] was used for the numerical implementation of the ordinal methods applied in our investigation.

### F. Machine learning artistic styles

We apply the extreme gradient boosting (XGBoost) [20] algorithm to classify the artistic styles in paintings using various sets of ordinal features. XGBoost is an ensemble method that incrementally combines decision trees to minimize a loss function augmented by a regularization term (a concept known as regularizing gradient boosting). This method incorporates a series of algorithmic, numerical, and computational optimizations, making this algorithm renowned for its high efficiency and exceptional performance, often ranking among the top solutions in machine learning competitions. To predict artistic styles, we select the 20 most common styles from our dataset, each represented by at least 1,500 artworks, totaling 100,707 images. We randomly split this dataset in a stratified manner, allocating 80% of images for training the XGBoost classifiers (with standard parameters) while the remaining is used for testing the predictions. We replicate these steps across ten independent realizations, calculating both the average and the standard deviation of the accuracy on the test sets.

We investigate eight sets of ordinal features for predicting artistic styles. Six sets use the ordinal encoding proposed by Bandt and Pompe [4], which does not explicitly account for identical values within the image partitions: the complexity-entropy plane ($H$ and $C$ values), the Fisher-Shannon plane ($H$ and $F$ values), the smoothness-structure plane ($\tau$ and $\kappa$ values), the probabilities of the three pattern types without ties ($q_1$, $q_2$, and $q_3$), the probabilities of the 24 possible patterns without ties (representing the standard ordinal distribution $P = \{p_i;\ i = 1, \ldots, n\}$ where identical values are handled based on their positions), and the same set of 24 possible patterns obtained resolving rank ties through the addition of small noise to partition values (this noise only disrupts equal-value degeneracy without affecting other ordering relationships). The remaining two feature sets, original to our work, explicitly consider the occurrence of identical values within the image partitions: the probability associated with each of the eleven pattern types ($q_A, q_B, q_C, \ldots,$ and $q_K$) and the probability of each of the 75 ordinal patterns. We further evaluate the average accuracy from two baseline classifiers: one stratified, which generates random predictions respecting the style frequency in the dataset, and one uniform, which assigns equal probability to each style.

Using the model trained on all 75 ordinal patterns, we assess the contribution of each pattern to the model's overall performance using permutation feature importance [21–23]. Originally proposed for random forests [21], this method has been extended into a model-agnostic framework [22] that consists of randomly shuffling the values of a single feature multiple times and measuring the decrease in the model's accuracy on the test set in each iteration. The premise of this technique is that disrupting the relationship between the shuffled feature and the style allows for the evaluation of the feature's significance based on the observed decline in accuracy; a greater decline indicates a higher importance of the feature. Figure 5C presents a box plot depicting the distribution of permutation importance values across 100 repetitions for each ordinal pattern, highlighting the top 10 patterns as ranked by their mean values. We rely on the scikit-learn [24] for the numerical implementation of this technique.

### G. Low-dimensional projection of the ordinal patterns probabilities

We use the uniform manifold approximation and projection (UMAP) dimensionality reduction algorithm [25] to project the 75-dimensional vectors representing the relative frequencies of ordinal patterns into a two-dimensional space. UMAP is considered one of the state-of-the-art dimensionality reduction techniques that relies on a weighted graph representation (fuzzy simplicial complex) obtained from a dissimilarity matrix (here calculated using Euclidean distance) determined from high-dimensional data points. This data is then projected into a lower-dimensional space (a plane in our case) using an approach that corresponds to a force-directed layout algorithm. Low-dimensional projections produced by UMAP have a degree of stochasticity inherent in the optimizing process, resulting in projections that are similar but not identical across runs of the same dataset. Consequently, UMAP projections are primarily used

for comparative and visualization purposes. We use the Python package *umap* [26] and set the number of neighbors to 150 to produce the visualization reported in Figure 2.

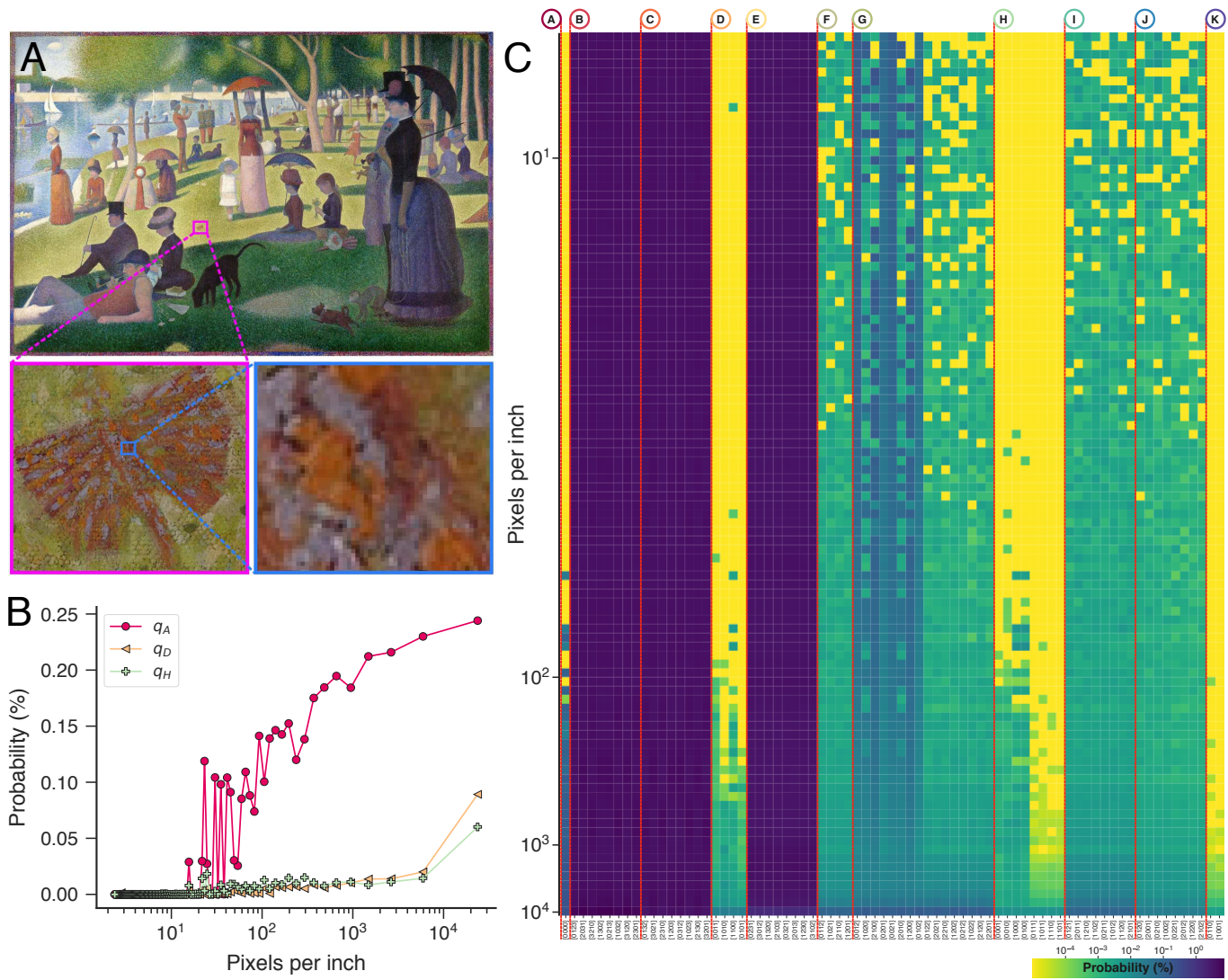

Figure S1. Effects of digital resolution on the prevalence of two-by-two ordinal patterns illustrated using Georges Seurat's "Sunday Afternoon on the Island of La Grande Jatte" (1884). (A) High-resolution digitalization of Seurat's painting, at 30,000×19,970 pixels, sourced from the Google Arts & Culture project. The physical canvas spans 308 cm in width and 208 cm in height, with the digital representation achieving a pixel density of nearly 24,000 pixels per inch (or over 9,000 pixels per square centimeter). Insets focus on a butterfly in the middle left of the painting, showcasing the intricacy of Seurat's technique. (B) Incidence of pattern types $A$ ($q_A$, circles), $D$ ($q_D$, triangles), and $H$ ($q_H$, crosses) as a function of the pixel density in downscaled versions of the original high-resolution image. These lower-resolution images are generated by sampling one pixel every $k$ pixels along both the horizontal and vertical axes for $k = 100, 99, \ldots, 1$, producing images ranging from 200×300 pixels (approximately one pixel per inch) up to the original resolution. (C) Probability of observing each of 75 ordinal patterns across varying resolutions in the downscaled images.

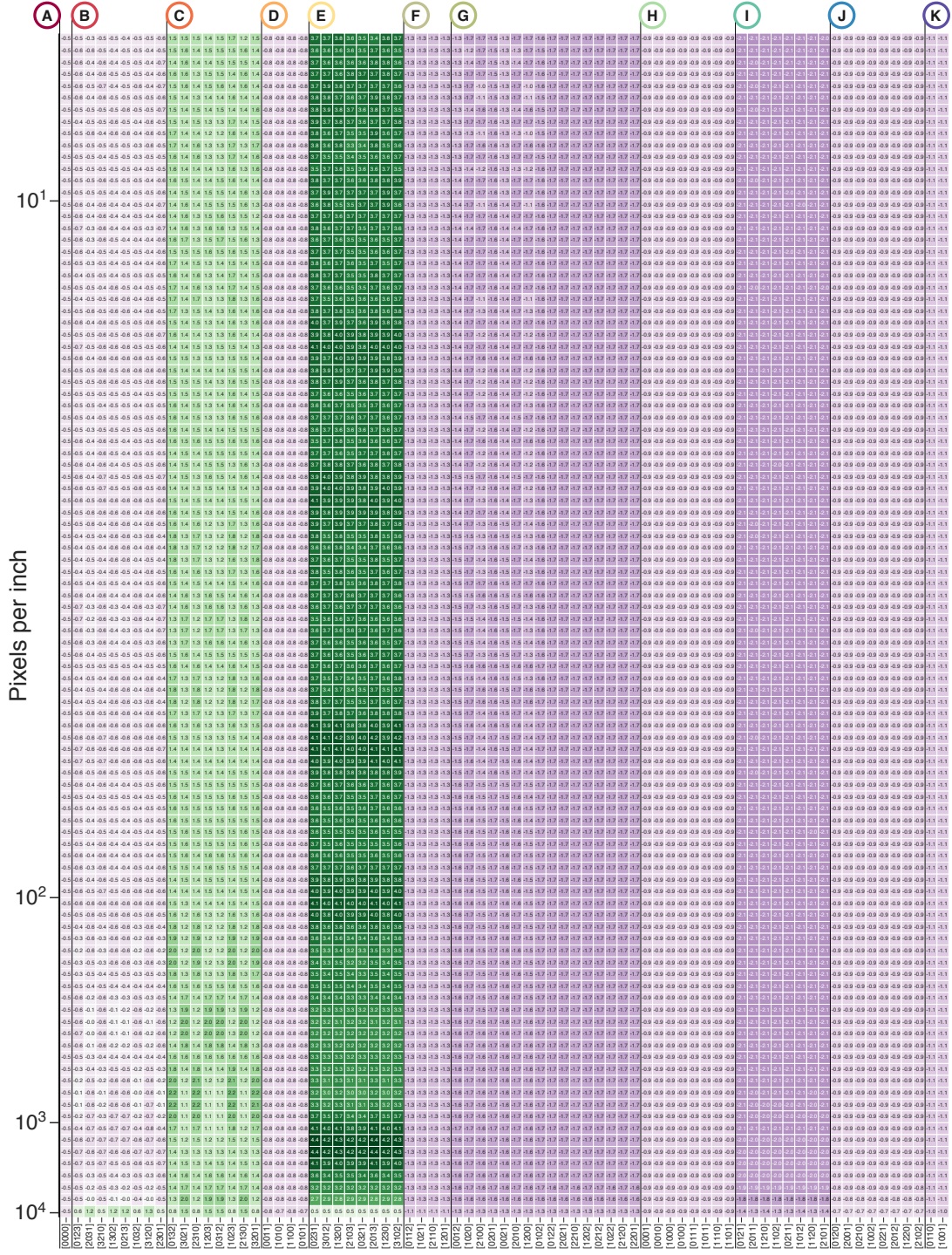

Figure S2. Effects of digital resolution on the standardized prevalence of two-by-two ordinal patterns illustrated using Georges Seurat's "Sunday Afternoon on the Island of La Grande Jatte" (1884). The lines in the matrix plot depict the $z$-score probability of each ordinal pattern calculated across varying resolutions in downscaled versions of the high-resolution digitalization of Seurat's painting. The $z$-score probability is calculated by subtracting the probability of observing an ordinal pattern at a given pixel density from the overall average probability across all images and dividing the result by the overall standard deviation of that pattern's probability across all images in our dataset. Positive values (represented by shades of green) indicate patterns occurring more frequently than the overall average, whereas negative values (represented by shades of purple) denote patterns occurring less frequently than the overall average, with deviations expressed in standard deviation units. The primary features of the ordinal distribution remain relatively stable across the pixel densities in lower-resolution versions.

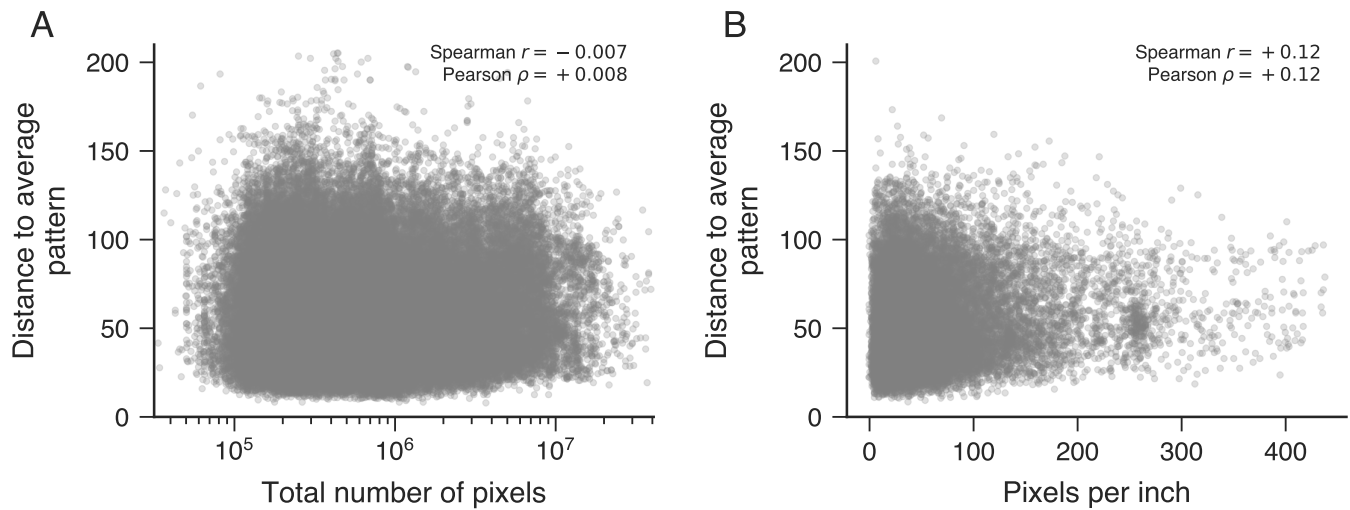

Figure S3. Absence of a clear association between the distance to the average ordinal pattern and image dimensions or resolutions. (A) Scatter plot of the distance to the average ordinal pattern versus the total number of pixels in all images in the dataset. (B) Scatter plot of the distance to the average ordinal pattern versus pixels per inch for all images containing information about the real dimensions of the artwork (approximately 20% of the dataset). Spearman's rank correlation ($r$) and Pearson's correlation ($\rho$) coefficients are shown within the panels. In both cases, there is no clear association between the distance to the average ordinal pattern and either the total number of pixels or pixels per inch, as evidenced by the very weak correlations observed.

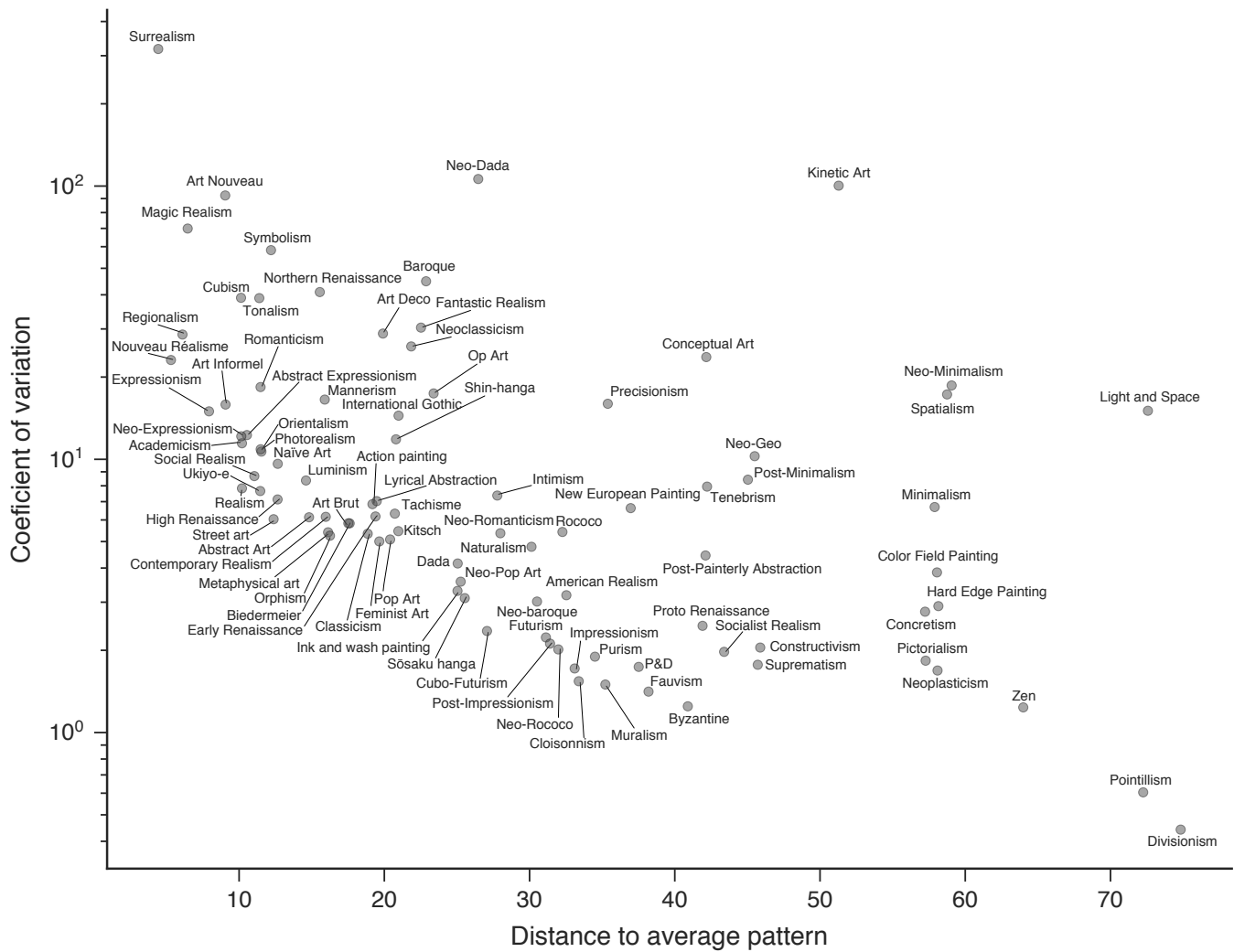

Figure S4. Negative association between the coefficients of variation of ordinal probabilities and the distance to the average pattern. The scatter plot shows the average absolute values of the coefficients of variation of the ordinal probabilities (in $z$-score units) versus the distance to the average pattern. The first quantity is calculated by estimating the coefficients of variation of the $z$-score probabilities of all ordinal patterns across all paintings within a given style, followed by averaging their absolute values. As detailed in the main text in the context of Figure 4, the distance to the average pattern is determined by the absolute sum of the $z$-score probabilities for each style. A general trend is observed, showing a decrease in the coefficient of variation as the distance to the average pattern increases.

File S1. The file `file_s1.csv` comprises a CSV file featuring 103 columns and 137,364 rows, excluding the header. Each row represents a painting. The initial four columns provide demographic details of the paintings (artist, artwork title, date of composition, and artistic style). Columns 5 to 78 contain the probabilities associated with each of the 75 ordinal patterns, arranged in the same sequence as presented in Figure 1 of the manuscript.



\end{document}